\documentclass[10pt,twocolumn,letterpaper]{article}
\usepackage[T1]{fontenc}
\usepackage[pagenumbers]{cvpr}
\definecolor{cvprblue}{rgb}{0.21,0.49,0.74}
\usepackage[breaklinks,colorlinks,allcolors=cvprblue,pdfusetitle]{hyperref}

\title{UniPlasmaTurb: A Unified Multi-Field Reconstruction and Forecasting Framework for Tokamak Edge Plasma Turbulence}
\author{%
  Xinxu Du\textsuperscript{1}\quad
  Xiao Wang*\textsuperscript{1}\quad
  Xiang Gao\textsuperscript{1}\quad
  Qingquan Yang*\textsuperscript{2}\\
  Yaowei Wang\textsuperscript{3,4}\quad
  Yonghong Tian\textsuperscript{3,5,6}\quad
  Jin Tang\textsuperscript{1}\quad
  Guosheng Xu\textsuperscript{2}\\[0.35em]
  {\normalsize \textsuperscript{1}\, School of Computer Science and Technology, Anhui University, Hefei 230601, China \\ 
               \textsuperscript{2}\, Institute of Plasma Physics, Chinese Academy of Sciences}\\
  {\normalsize \textsuperscript{3}\, Peng Cheng Laboratory, Shenzhen, China \\ 
               \textsuperscript{4}\, Harbin Institute of Technology (HITSZ), Shenzhen, China \\ 
               \textsuperscript{5}\, School of Computer Science, Peking University, China \\ 
               \textsuperscript{6}\, School of Electronic and Computer Engineering, Shenzhen Graduate School, Peking University, China} 
}
\hypersetup{pdfauthor={Xinxu Du, Xiao Wang, Xiang Gao, Qingquan Yang, Yaowei Wang, Yonghong Tian, Jin Tang, Guosheng Xu}}

\makeatletter
\renewcommand{\@maketitle}{%
  \begingroup
    \centering
    {\Large\bfseries \@title\par}
    \medskip
    {\large \@author\par}
    \medskip
  \endgroup
}
\newenvironment{suppanchoredtable}{\par\begingroup\def\@captype{table}\captionsetup{hypcap=false}\centering\small}{\par\endgroup}
\newcommand{\suppresetcolumns}{\global\@colht\textheight\global\@colroom\textheight\global\vsize\textheight}
\makeatother

\begin{document}
\maketitle

\begin{abstract}
Observations of tokamak edge turbulence have limited resolution, sparse sampling, or missing physical fields.
Reconstruction and forecasting must accommodate these conditions while retaining localized density structures that average field errors can obscure.
This work constructs a coupled model of pressure, density, vorticity, and effective potential, including background-gradient driving and scrape-off-layer losses, and introduces UniPlasmaTurb for unified super-resolution, field completion, and forecasting.
Observation histories, masks, and task indicators enter a shared network. Output projections preserve reconstruction measurements.
Region-balanced supervision constrains the amplitude and local spatial coverage of strong positive density fluctuations while suppressing spurious background responses.
Predicted four-field states are fed back for autoregressive forecasting.
On simulated plasma trajectories, UniPlasmaTurb achieves lower reconstruction errors than the evaluated baselines and higher strong-density-structure recall at all evaluated forecasting horizons, with the lowest field error at the shortest horizon.
Different rankings by field error and structure recall show the need to evaluate both overall accuracy and localized structures.
The dataset and code will be publicly released on \url{https://github.com/Event-AHU/OpenFusion} 
\end{abstract}

\section{Introduction}
\label{sec:introduction}
Images and videos record physical processes such as motion, deformation, and flow.
Reconstruction and forecasting link computer vision with physical science.
Recent CVPR work estimates dynamical parameters from video in Learning Physics From Video~\cite{garcia2025learningphysics}, reconstructs and forecasts three-dimensional fluids in FluidNexus~\cite{gao2025fluidnexus}, and combines data consistency with physical cues for fluid super-resolution in ReMD~\cite{li2026remd}.
These studies extend visual learning to spatially and temporally coupled scientific data in AI for Science.

\begin{figure*}[!t]
\centering
\includegraphics[width=0.98\textwidth]{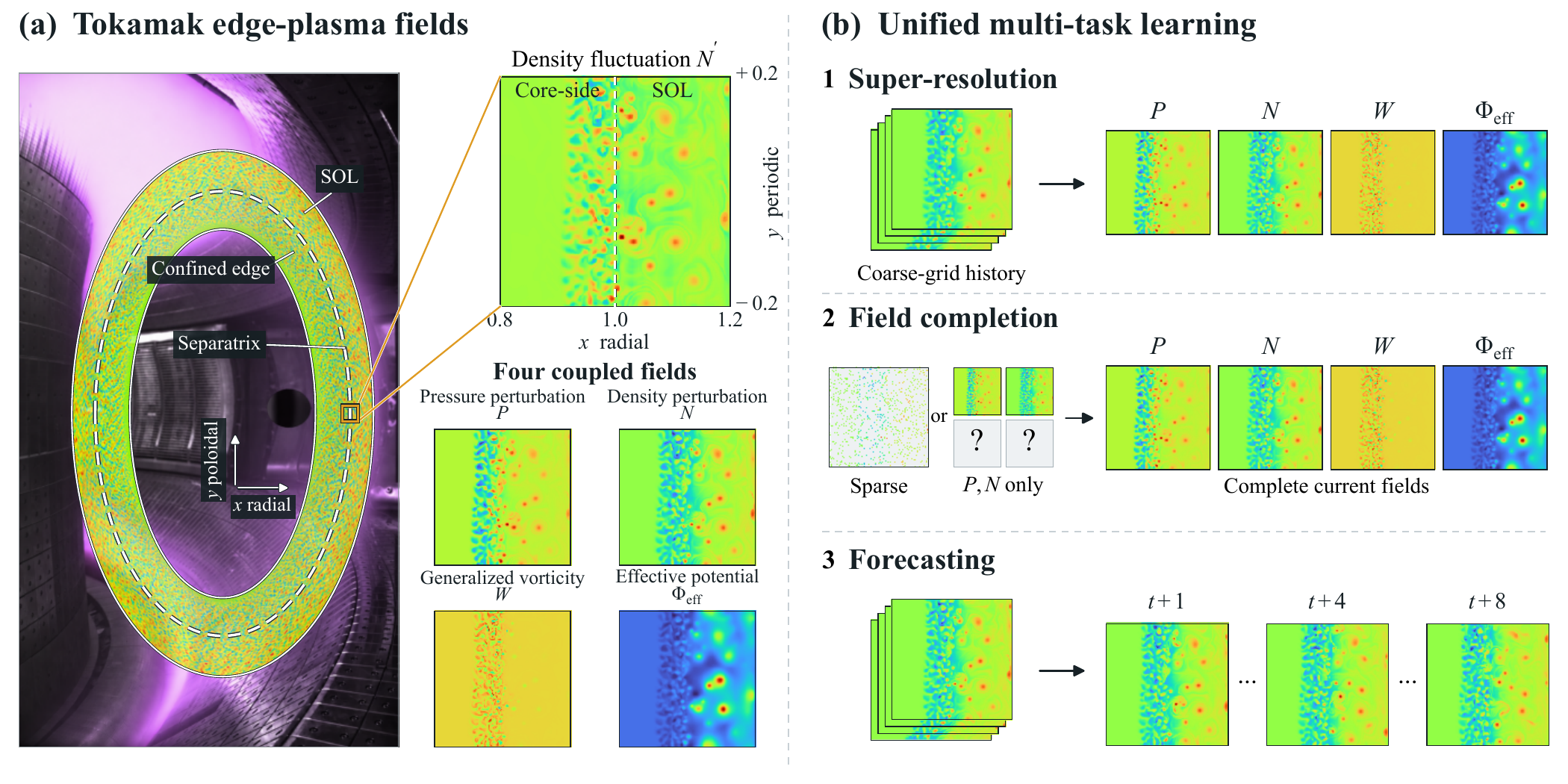}
\caption{\textbf{Tokamak edge turbulence and unified learning tasks.}
(a) Local density fluctuations and scrape-off-layer (SOL) transport~\cite{dippolito2011convective}, contextualized by global edge simulations~\cite{stegmeir2019global} and X-point studies~\cite{galassi2019tokamak}. The annulus is a local-field illustration, not a global simulation. The device overlay is not to scale. The box marks a 4\,cm radial window. The adjacent $N'=N-\langle N\rangle_y$ uses the periodic-direction mean and has its SOL interface at $x=1$.
(b) Observation histories and targets from one training trajectory.}
\label{fig:physical_overview}
\end{figure*}

Tokamak edge turbulence poses related problems in nuclear fusion: intermittent density enhancements observed in experiments~\cite{zweben2007edge} and fluid simulations~\cite{garcia2004computations} carry particles across magnetic field lines, making their strength, distribution, and evolution important for transport analysis~\cite{dippolito2011convective}.
This work studies four coupled simulated fields---pressure, density, vorticity, and effective potential---through super-resolution (SR), field completion (FC), and forecasting (FP), as illustrated in Fig.~\ref{fig:physical_overview}.
The central challenge is to enable one model to recover unknown states under different observation conditions while preserving measurements and localized strong density structures.
Low average field error can conceal missed structures. Strengthening local responses must also suppress spurious background peaks.

Physical-field learning has advanced reconstruction, forecasting, and small-scale recovery, but three issues remain for the four-field tasks considered here:
\begin{itemize}
\item \textbf{Incomplete task coverage.} Plasma flow matching supports snapshot generation, SR, and sparse completion~\cite{guo2026generative}, but forecasting and missing-field completion remain unvalidated.
\item \textbf{No guarantee of observation consistency.} GLU unifies sparse reconstruction and forecasting~\cite{wang2026glu}, but its learned decoder does not guarantee exact measurement preservation. SR and FC require downsampling and observed-value consistency, respectively.
\item \textbf{Insufficient local structure supervision.} Fluctuation-statistics and spectral losses improve turbulence detail~\cite{balasubramanian2026sharper}, but do not determine strong structures' locations or coverage. Amplitude errors, missed structures, and false background responses require direct supervision.
\end{itemize}

To address these issues, this work constructs a coupled four-field edge-turbulence simulation system and dataset and introduces \textbf{UniPlasmaTurb}, a unified reconstruction and forecasting framework.
Historical fields, masks, and task indicators let one shared network reconstruct current states or predict future states.
Output projections enforce downsampling consistency for SR and preserve measured values for FC. Forecasting feeds predicted four-field states back into the history.
Region-balanced supervision fits strong density amplitudes and penalizes spurious background peaks, while local soft occupancy constrains spatial coverage.
Both losses average structure and background errors separately to prevent the larger background from dominating supervision.
The main contributions are:
\begin{itemize}
\item \begin{samepage}\textbf{A coupled simulation system, dataset, and evaluation framework.} Varying initial perturbations generates four-field trajectories for all three tasks, evaluated by field error, structure recall, and false discovery rate.
\par
\end{samepage}
\item \textbf{Unified inference with observation consistency.} One shared model supports reconstruction and forecasting, with explicit output constraints that preserve reconstruction measurements.
\item \textbf{Region-balanced structure supervision.} Separately normalized amplitude and coverage losses constrain strong density structures and suppress false background responses.
\end{itemize}

\section{Related Work}
\label{sec:related_work}
\paragraph{Spatiotemporal modeling of physical systems.}
Data-driven fluid mechanics~\cite{brunton2020machine} and turbulence modeling~\cite{duraisamy2019turbulence} study spatially and temporally coupled fields.
Neural operators map between function spaces~\cite{kovachki2023neural}.
The Fourier neural operator (FNO) learns Fourier-space integral kernels~\cite{li2021fourier}. The convolutional neural operator (CNO) links continuous representations and discrete computation through convolution and resampling~\cite{raonic2023cno}. PDE-Transformer models cross-field interactions through attention~\cite{holzschuh2025pdetransformer}.
For plasma turbulence, multichannel flow matching learns joint density--potential distributions for snapshot generation and reconstruction~\cite{guo2026generative}.
This work combines spatial reconstruction, cross-field completion, and forecasting in a coupled four-field dataset and framework.

\paragraph{Unified inference under observation constraints.}
Shared representations support multitask learning~\cite{caruana1997multitask}, while observation models define inverse problems~\cite{stuart2010inverse}.
RAM conditions on observation operators and noise to solve multiple imaging inverse problems~\cite{supp_terris2025ram}. GLU combines global, local, and uncertainty representations for reconstruction and forecasting~\cite{wang2026glu}. OmniField uses modality-conditioned continuous fields for reconstruction, forecasting, and missing-modality inference~\cite{valencia2026omnifield}.
Null-space networks explicitly preserve measurements by restricting learned corrections~\cite{schwab2019nullspace}.
UniPlasmaTurb combines a shared observation interface with task-specific output projections for measurement-consistent reconstruction.

\paragraph{Local structures and region-balanced supervision.}
Neural spectral bias favors lower frequencies~\cite{rahaman2019spectral}.
Focal Loss reduces easy-background dominance in dense detection~\cite{lin2017focal}.
HiNOTE learns pixel contributions from a spectral-resizing prior~\cite{luo2024hinote}. IRNO combines residual correction with progressive spectral loss~\cite{liu2026irno}. Turbulence losses based on fluctuation statistics and spectra improve small-scale recovery~\cite{balasubramanian2026sharper}.
This work directly supervises strong-density amplitude and local coverage, normalizing structure and background losses separately.

\section{Coupled Edge-Turbulence Equations And Learning Tasks}
\label{sec:problem}
\subsection{Physical system and simulated data}
\label{sec:physical_model}
The state $U=(P,N,W,\Phi_{\mathrm{eff}})$ comprises pressure perturbation $P$, density perturbation $N$, generalized vorticity $W$, and effective potential $\Phi_{\mathrm{eff}}$.
The perturbations are relative to fixed backgrounds $p_0(x)$ and $n_0(x)$.

The local domain has radial coordinate $x$ and periodic coordinate $y$. The separatrix divides the confined edge and scrape-off layer (SOL).
Fields are periodic in $y$ and zero at radial boundaries, with
\begin{equation}
\begin{aligned}
\partial_t Q+\mathbf u\cdot\nabla Q&=D\Delta Q+S_Q,\quad Q\in\{P,N,W\},\\
\nabla\cdot(C_\phi\nabla\Phi_{\mathrm{eff}})&=a^2W.
\end{aligned}
\label{eq:physical_system}
\end{equation}
Here $D$ is diffusivity, $a$ a spatial scale factor, and $C_\phi(x)$ depends on background density.
For $(Q,q_0)=(P,p_0)$ or $(N,n_0)$, the source is
\begin{equation}
S_Q=-u_x\partial_xq_0-\chi_{\mathrm{SOL}}Q/\tau_p,
\label{eq:pressure_density_sources}
\end{equation}
where $\chi_{\mathrm{SOL}}$ indicates the SOL and $\tau_p$ is its loss timescale.
The two terms describe radial drift across background gradients and perturbation losses in the SOL.
$S_W$ combines pressure-gradient curvature driving with region-specific current terms depending on density and potential.
The drift $\mathbf u=(E_y,-E_x)/(aB_\phi)$ follows
\begin{equation}
E_i=-\frac{T_{\mathrm{ref}}}{a}\left(\partial_i\Phi_{\mathrm{eff}}-\frac{\partial_iP}{n_0}\right),\quad i\in\{x,y\},
\label{eq:electric_field_drift}
\end{equation}
with reference temperature $T_{\mathrm{ref}}$ and toroidal field $B_\phi$.
Thus pressure and potential determine transport, while vorticity and potential also satisfy a spatial constraint, motivating joint four-field inference.
Trajectories vary the initial pressure perturbation's amplitude, location, width, spectrum, and phase while keeping backgrounds and boundaries fixed.
Appendix~\ref{supp:physics} gives complete sources, profiles, and numerical generation details.

\subsection{Observation models and prediction targets}
\label{sec:learning_tasks}
Histories ending at $t$ support current-state reconstruction or future-state prediction:
\begin{itemize}
\item \textbf{Super-resolution (SR).} Low-pass downsampling $R$ produces $Y_t=RU_t$.
The model uses coarse-grid histories to recover the current fine-scale state while satisfying $R\widehat U_t=Y_t$.
\item \textbf{Field completion (FC).} Sparse spatial completion ($\mathrm{FC_s}$) uses a spatial mask shared across fields and historical frames. Missing-field completion ($\mathrm{FC_c}$) observes complete channels and infers the others.
Mask $M$ equals one on observed entries and zero elsewhere. Observations are $Y_t=M\odot U_t$, and completion must satisfy $M\odot\widehat U_t=Y_t$.
\item \textbf{Temporal forecasting (FP).} Complete histories predict the next four-field state, which feeds back for multi-step prediction.
Besides field accuracy, evaluation measures strong positive density fluctuations $N'=N-\langle N\rangle_y$, where $\langle\cdot\rangle_y$ is the periodic-direction mean at each radial location.
\end{itemize}
Observation operators and structure metrics are detailed in Appendix~\ref{supp:protocol}.

\begin{figure*}[!t]
\centering
\includegraphics[width=0.97\textwidth]{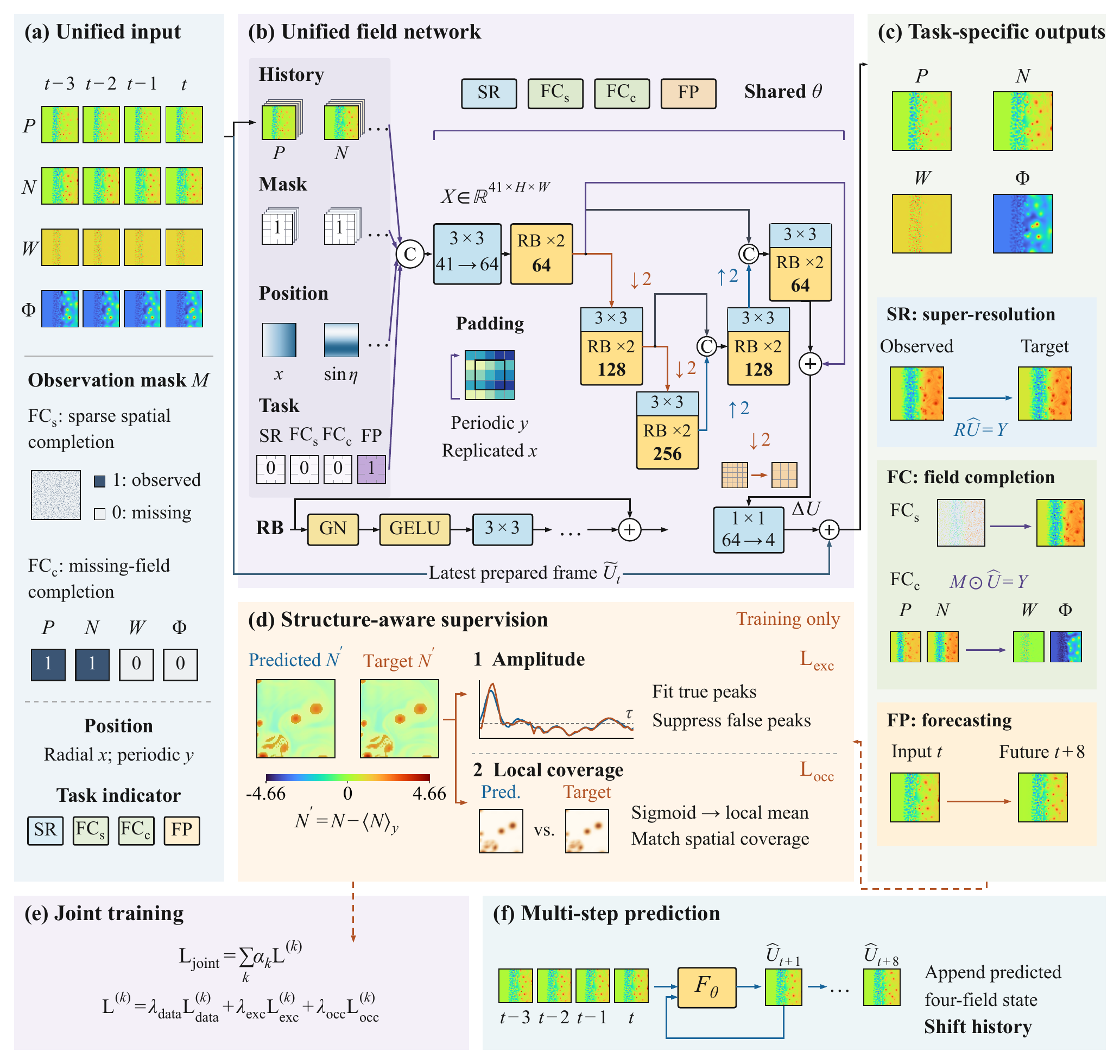}
\caption{\textbf{UniPlasmaTurb: unified multi-field reconstruction and forecasting.}
(a,b) Fields, masks, positions, and task indicators enter a shared residual U-Net. (c,f) Projections preserve reconstruction measurements. Predicted states feed back for forecasting. (d,e) Region-balanced amplitude and soft-occupancy losses fit strong structures, suppress background exceedances, and constrain coverage. C, RB, and GN denote concatenation, residual block, and group normalization. $\Phi$ denotes effective potential. SR/FC show observations and simulation targets. Crops illustrate full-SOL supervision, with true structures shaded orange.}
\label{fig:uniplasmaturb_framework}
\end{figure*}

\section{UniPlasmaTurb Framework}
\label{sec:method}
UniPlasmaTurb combines a shared network, observation-consistent outputs, and region-balanced supervision (Fig.~\ref{fig:uniplasmaturb_framework}).
The input distinguishes observed information from unknown content, while training addresses both overall field accuracy and local structure recovery.

\subsection{Unified input and shared network}
\label{sec:method_backbone}
For history length $L$ and task $k\in\mathcal K=\{\mathrm{SR},\mathrm{FC_s},\mathrm{FC_c},\mathrm{FP}\}$, inputs are mapped to a common target grid: SR uses periodic bilinear interpolation, FC zero-fills missing entries, and FP uses complete fields.
Processed histories $\widetilde U_{t-L+1:t}^{(k)}$ and availability masks $\widetilde M_{t-L+1:t}^{(k)}$ form
\begin{equation}
X_t^{(k)}=\operatorname{Concat}\!\left(\widetilde U_{t-L+1:t}^{(k)},\widetilde M_{t-L+1:t}^{(k)},G,c_k\right),
\label{eq:uniplasmaturb_input}
\end{equation}
where concatenation is channelwise, $G$ encodes radial position, sine/cosine of periodic position, and grid scales, and $c_k$ is a spatially broadcast task indicator.
SR's interpolated input is fully available. Output projection retains the original coarse observations.

A shared adapter feeds a residual U-Net~\cite{ronneberger2015unet}.
The encoder extracts multiscale features. The decoder concatenates same-scale encoder features to recover fine spatial information, with an additional finest-scale feature residual.
Residual blocks~\cite{he2016deep} contain two sequences of group normalization~\cite{wu2018group}, GELU~\cite{hendrycks2016gelu}, and convolution, connected by an identity shortcut.
Convolutions use periodic $y$ padding and replicated radial boundaries. Low-pass downsampling retains radial endpoints, and bilinear upsampling connects the periodic endpoints.
A pointwise head predicts a four-field increment:
\begin{equation}
Z^{(k)}=\widetilde U_t^{(k)}+\Delta_\theta(X_t^{(k)}).
\label{eq:uniplasmaturb_residual_output}
\end{equation}
The adapter, backbone, and head share parameters $\theta$ across tasks.
For SR/FC, the increment corrects the latest observation toward the complete current state. For FP, it advances that state to the next time step.
Architecture details are in Appendix~\ref{supp:architecture}.

\subsection{Task-specific observation consistency}
\label{sec:method_execution}
Final outputs apply the corresponding observation constraint:
\begin{equation}
\widehat U^{(k)}=
\begin{cases}
Z^{(k)}+R^\dagger(Y_t-RZ^{(k)}),&k=\mathrm{SR},\\
M\odot Y_t+(1-M)\odot Z^{(k)},&k\in\{\mathrm{FC_s},\mathrm{FC_c}\},\\
Z^{(k)},&k=\mathrm{FP}.
\end{cases}
\label{eq:uniplasmaturb_projection}
\end{equation}
The fixed right inverse $R^\dagger$ is constructed from the actual downsampling operator and satisfies $RR^\dagger=I$.
It maps the coarse-grid measurement residual back to the fine grid, giving $R\widehat U=Y_t$ to numerical precision.
FC copies observed values and predicts missing entries. FP has no future measurements to impose.
Training and inference use these same final outputs. The right-inverse construction is in Appendix~\ref{supp:projection}.

\subsection{Region-balanced structure supervision}
\label{sec:method_structure}
Strong density structures occupy only part of the SOL.
Their amplitude and local coverage receive separate supervision, with independent normalization over structure and background regions.

For predicted and true density fluctuations $p,n$, define $A=\{i\in\Omega_{\mathrm{SOL}}:n_i>\tau\}$ using a training-derived threshold $\tau$, and the regional mean
\begin{equation}
\mathcal B_B(e)=\frac{\sum_{i\in B}e_i}{\max(|B|,1)}.
\label{eq:uniplasmaturb_balanced_mean}
\end{equation}
Complements are relative to the SOL. Empty regions contribute zero.
\paragraph{Amplitude and false-peak supervision.}
The exceedance loss is
\begin{equation}
\mathcal L_{\mathrm{exc}}=\tfrac12\left[\mathcal B_A((p-n)^2)+\mathcal B_{A^c}([p-\tau]_+^2)\right],
\label{eq:uniplasmaturb_exceedance}
\end{equation}
where $[z]_+=\max(z,0)$.
The terms fit true strong amplitudes and penalize background exceedances. Separate normalization prevents the larger background from dominating through pixel count. Data loss constrains weaker and negative fluctuations.
\paragraph{Local coverage through soft occupancy.}
Soft occupancy and structure neighborhoods are
\begin{equation}
o(v)=\mathcal A_r\!\left[\sigma\!\left(\frac{v-\tau}{T}\right)\right],\quad S=\{i:(\mathcal A_r\mathbf1_A)_i>0\},
\label{eq:uniplasmaturb_occupancy_map}
\end{equation}
where $\sigma$ is the sigmoid function, $T$ controls smoothness, and $\mathcal A_r$ averages $r\times r$ neighborhoods with periodic padding in $y$ and replication at the SOL radial edges.
$S$ contains positions whose neighborhoods intersect true strong structures.
For $e_o=(o(p)-o(n))^2$, the coverage loss is
\begin{equation}
\mathcal L_{\mathrm{occ}}=\tfrac12[\mathcal B_S(e_o)+\mathcal B_{S^c}(e_o)].
\label{eq:uniplasmaturb_occupancy_loss}
\end{equation}
It balances coverage errors near true structures and in the background, complementing amplitude supervision over the entire SOL.

\subsection{Joint training and multi-step forecasting}
\label{sec:method_training}
Data loss is mean squared error (MSE) over complete SR/FP outputs and unobserved FC entries. The joint objective is
\begin{equation}
\mathcal L_{\mathrm{joint}}=\sum_{k\in\mathcal K}\alpha_k\sum_{j\in\{\mathrm{data},\mathrm{exc},\mathrm{occ}\}}\lambda_j\mathcal L_j^{(k)},
\label{eq:uniplasmaturb_joint_loss}
\end{equation}
with task weights $\alpha_k$ and loss weights $\lambda_j$.
Each update constructs all task observations from the same histories and accumulates their weighted gradients.
SR/FC target the last historical state, while FP targets the next state.
When FC observes density completely, projection makes both structure losses zero, leaving data error on the missing channels.

Let $F_\theta$ include input processing and output rules, with $\mathcal H_0=(U_{t-L+1},\ldots,U_t)$.
For $h\ge0$, autoregressive forecasting follows
\begin{equation}
\widehat U_{t+h+1}=F_\theta(\mathcal H_h;\mathrm{FP}),\quad
\mathcal H_{h+1}=\operatorname{shift}(\mathcal H_h,\widehat U_{t+h+1}),
\label{eq:uniplasmaturb_rollout}
\end{equation}
where shift removes the oldest state and appends the complete four-field prediction.

\begin{table*}[!t]
\centering
\caption{\textbf{Multi-step forecasting.} Entries report four-field MSE and structure recall $R$ (\%), with best values in bold. Parameters are in millions and cumulative training time in GPU-hours. DLL-NO uses the mean of 32 sampled trajectories.}
\label{tab:fp_main}
\small
\setlength{\tabcolsep}{3pt}
\resizebox{0.95\textwidth}{!}{%
\begin{tabular}{@{}llcccrr@{}}
\toprule
Method & Venue & \shortstack{H8\\MSE$\downarrow$ / R$\uparrow$}
       & \shortstack{H16\\MSE$\downarrow$ / R$\uparrow$}
       & \shortstack{H32\\MSE$\downarrow$ / R$\uparrow$}
       & \shortstack{Params.\\(M)} & \shortstack{GPU time\\(GPU-h)} \\
\midrule
FNO~\cite{li2021fourier} & ICLR 2021 & 0.7693 / 37.98 & 0.9998 / 33.04 & 1.2129 / 27.35 & 4.755 & 29.55 \\
CNO~\cite{raonic2023cno} & NeurIPS 2023 & 0.6093 / 46.79 & 1.1130 / 34.02 & 2.0839 / 24.97 & 4.888 & 29.29 \\
PDE-Transformer~\cite{holzschuh2025pdetransformer} & ICML 2025 & 0.5195 / 53.21 & 0.8152 / 35.33 & 1.0217 / 30.59 & 4.744 & 36.63 \\
LinearNO~\cite{hu2025linearno} & AAAI 2026 & 0.6692 / 5.11 & 0.7575 / 3.26 & \textbf{0.8500} / 4.11 & 4.747 & 49.16 \\
CFO-U-Net~\cite{hou2026cfo} & ICLR 2026 & 0.8586 / 34.60 & 0.9756 / 34.78 & 1.0144 / 33.41 & 8.097 & 22.56 \\
DLL-NO~\cite{park2026dll} & ICML 2026 & 0.4621 / 19.91 & \textbf{0.5562} / 0.65 & $6.9241\times10^3$ / 0.54 & 55.771 & 23.84 \\
\midrule
UniPlasmaTurb & This work & \textbf{0.4566} / \textbf{73.99} & 0.8310 / \textbf{47.39} & 1.1103 / \textbf{38.70} & 4.788 & 30.78 \\
\bottomrule
\end{tabular}
}
\end{table*}

\begin{figure*}[!t]
\centering
\includegraphics[width=0.98\textwidth]{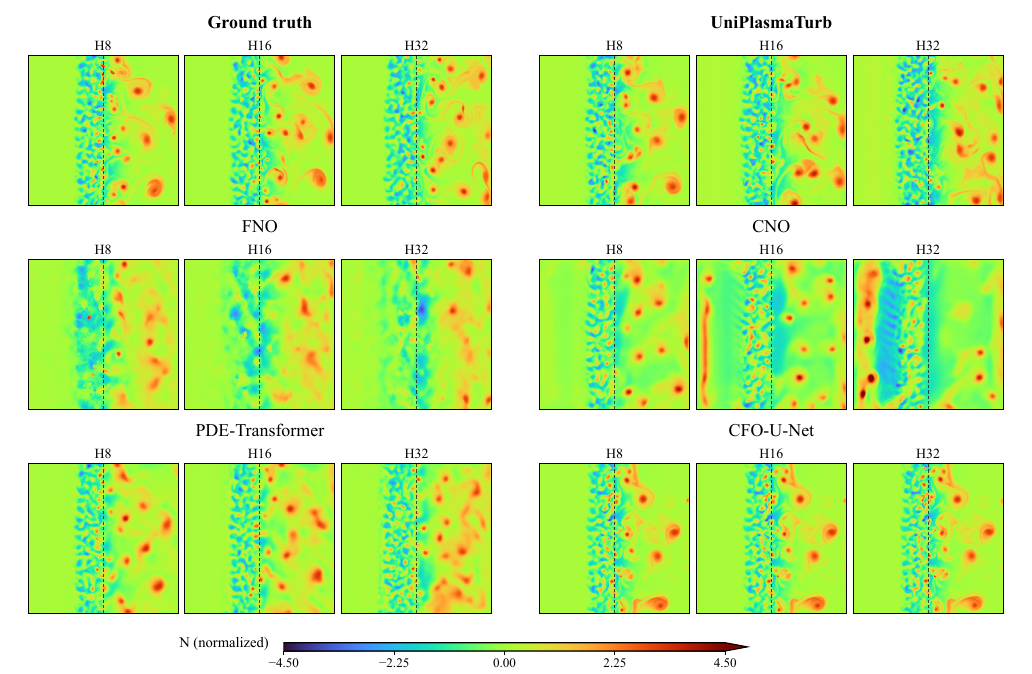}
\caption{\textbf{Density forecasts at H8, H16, and H32.} All panels use the full domain and common normalized scale $[-4.5,4.5]$. The color-bar extension marks larger values. Dashed lines indicate the SOL boundary.}
\label{fig:forecast_qualitative}
\end{figure*}

\section{Experiments}
\label{sec:experiments}
\subsection{Experimental setup}
\paragraph{Data and tasks.}
The 144 trajectories are split by initial-condition group into 96 training, 24 validation, and 24 test trajectories, keeping phase realizations of each group together.
UniPlasmaTurb trains on one predetermined trajectory per training group (24 total), also used for input normalization. Reporting scales and structure thresholds use all 96 training trajectories.
Four historical states, separated by $1/3\,\mu\mathrm{s}$, are represented on a $256\times257$ grid.
SR observes $128\times129$ fields, $\mathrm{FC_s}$ observes about 10\% of locations, and $\mathrm{FC_c}$ observes $P,N$ and recovers $W,\Phi_{\mathrm{eff}}$.
All comparisons and analyses use three fixed windows per test trajectory (72 total), with FP evaluated at H8/H16/H32. Appendix~\ref{supp:protocol} specifies splits, normalization, and windows.

\paragraph{Training and evaluation.}
The network has base width 64 and 4.79M parameters, trained for 32 epochs (3,552 updates) with AdamW~\cite{loshchilov2019decoupled}, batch size 128, and weight decay 0.01.
The learning rate warms from $10^{-4}$ to $10^{-3}$ over five epochs and decays to $10^{-4}$ by cosine scheduling.
Task weights for SR, $\mathrm{FC_s}$, $\mathrm{FC_c}$, and FP are $(1/3,1/6,1/6,1/3)$. Data/amplitude/occupancy weights are $(1.1,0.1,0.05)$, with a $7\times7$ occupancy neighborhood.
All tasks use epoch-32 parameters. Baselines retain their established training versions (Appendix~\ref{supp:baselines}).
Hardware comprises NVIDIA A800-SXM4-80GB GPUs and dual Intel Xeon Gold 6348 processors.

SR/FP report four-field MSE, while FC scores unobserved entries.
Strong structures exceed the training-SOL 95th percentile of $|N'|$, with eight-connectivity, area at least 36 pixels, and periodic matching within 16 pixels.
Recall $R$ pools matched and true structure counts over test windows.
Appendix~\ref{supp:matching} defines matching and false discovery rate (FDR). Appendix~\ref{supp:comparisons} reports complete metrics.
UniPlasmaTurb's reported GPU time covers all three tasks.

\begin{table*}[!t]
\centering
\caption{\textbf{Reconstruction MSE.} SR uses complete fields, and FC uses unobserved entries. Costs follow Table~\ref{tab:fp_main}. Dashes denote unevaluated settings. Best values are bold.}
\label{tab:restoration_main}
\small
\setlength{\tabcolsep}{5pt}
\begin{tabular*}{\textwidth}{@{\extracolsep{\fill}}llrrrrr@{}}
\toprule
Method & Venue & SR$\downarrow$ & $\mathrm{FC_s}\downarrow$ & $\mathrm{FC_c}\downarrow$ & \shortstack{Params.\\(M)} & \shortstack{GPU time\\(GPU-h)} \\
\midrule
BasicVSR++~\cite{chan2022basicvsrpp} & CVPR 2022 & 0.00821 & --- & --- & 7.195 & 40.98 \\
HiNOTE~\cite{luo2024hinote} & ICML 2024 & 0.01824 & --- & --- & 4.715 & 20.37 \\
UCAN~\cite{tan2026ucan} & CVPR 2026 & 0.00845 & --- & --- & 0.692 & 27.10 \\
IRNO~\cite{liu2026irno} & ICML 2026 & 0.00830 & --- & --- & 6.919 & 37.60 \\
RAM~\cite{supp_terris2025ram} & ICLR 2026 & 0.01983 & 0.08109 & 0.05105 & 8.750 & 106.82 \\
4DVarNet~\cite{beauchamp2023varnet} & GMD 2023 & --- & 0.08111 & 0.06595 & 0.169 & 28.63 \\
FLRONet~\cite{dang2025flronet} & JCISE 2026 & --- & 0.21481 & --- & 4.847 & 21.93 \\
LANO~\cite{hou2026lano} & AAAI 2026 & --- & 0.14693 & --- & 11.242 & 20.28 \\
PhySense~\cite{ma2025physense} & NeurIPS 2025 & --- & 0.68471 & --- & 12.762 & 23.03 \\
OmniField~\cite{valencia2026omnifield} & ICLR 2026 & --- & --- & 0.97527 & 41.775 & 30.12 \\
\midrule
UniPlasmaTurb & This work & \textbf{0.00813} & \textbf{0.04147} & \textbf{0.02042} & 4.788 & 30.78 \\
\bottomrule
\end{tabular*}
\end{table*}

\begin{figure*}[!t]
\centering
\includegraphics[width=0.96\textwidth]{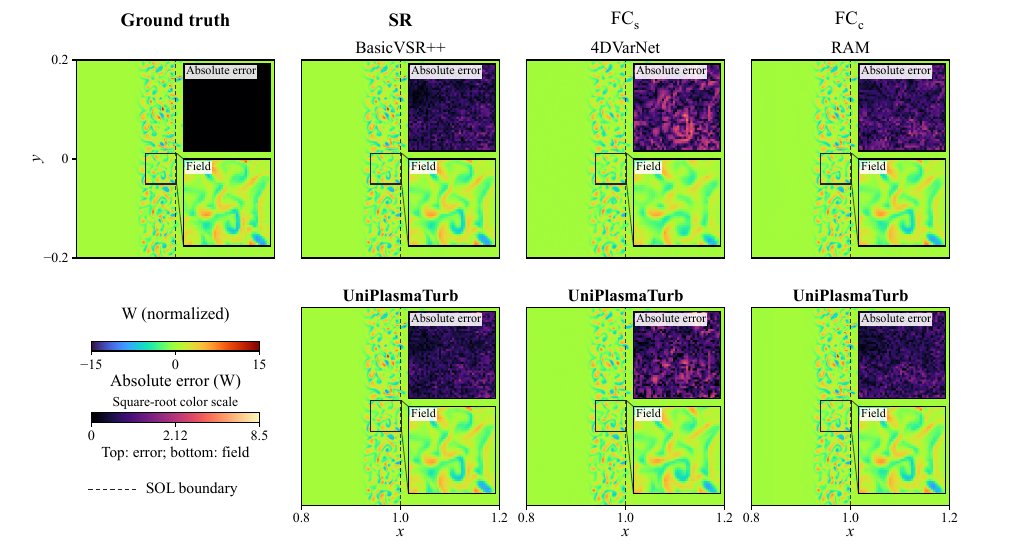}
\caption{\textbf{Vorticity reconstruction.} Each task compares UniPlasmaTurb with its indicated baseline at the same target frame and region. Upper insets show absolute errors, and lower insets show local fields. Field and error scales are shared. Error colors use a square-root scale with ticks in actual units.}
\label{fig:restoration_qualitative}
\end{figure*}

\subsection{Multi-step forecasting}
UniPlasmaTurb has the highest structure recall at H8/H16/H32 and the lowest field MSE at H8 (Table~\ref{tab:fp_main}).
Recall reaches 73.99\%, 47.39\%, and 38.70\%, exceeding the strongest baseline at each horizon by 20.78, 12.07, and 5.30 percentage points, respectively.
At H32, LinearNO has lower MSE (0.8500) but only 4.11\% recall, showing that low average error can conceal missed structures.
DLL-NO's large H32 MSE comes from severe long-horizon instability in a few windows.
Figure~\ref{fig:forecast_qualitative} shows UniPlasmaTurb retaining localized strong responses and fine SOL-boundary variations, while CFO-U-Net changes little across horizons. Position and shape deviations increase at longer horizons.
Appendix~\ref{supp:field_forecast} provides fieldwise results, FDR, and additional qualitative comparisons.

\begin{table}[!t]
\centering
\caption{\textbf{Ablation structure recall (\%).} All variants use epoch 32. Recovery learning-rate schedules differ. Appendix~\ref{supp:ablation_settings} gives configurations and complete metrics.}
\label{tab:structure_evidence}
\small
\setlength{\tabcolsep}{3pt}
\begin{tabular}{@{}lrrr@{}}
\toprule
Variant & H8$\uparrow$ & H16$\uparrow$ & H32$\uparrow$ \\
\midrule
Full model & \textbf{73.99} & \textbf{47.39} & \textbf{38.70} \\
No structure supervision & 66.05 & 42.93 & 31.35 \\
No regional balancing & 68.12 & 43.26 & 36.86 \\
Replicate convolution padding & 65.94 & 39.35 & 34.27 \\
\bottomrule
\end{tabular}
\end{table}

\begin{figure}[t]
\centering
\includegraphics[width=\linewidth]{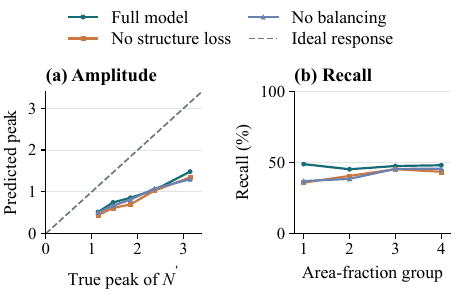}
\caption{\textbf{Structure supervision at H16.} (a) Mean predicted peaks within all true structures, grouped by true peak strength. The dashed line marks the ideal response. (b) Recall by true SOL structure area fraction. Groups use training-set boundaries (Appendix~\ref{supp:area_fraction}).}
\label{fig:structure_supervision_analysis}
\end{figure}

\begin{figure}[t]
\centering
\includegraphics[width=\linewidth]{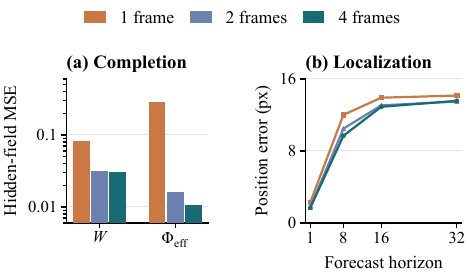}
\caption{\textbf{Historical information with fixed weights.} (a) Missing-field MSE (log scale). (b) Periodic localization error with a 16-pixel penalty per missed structure. Distinct states fill four slots by repetition.}
\label{fig:history_information_analysis}
\end{figure}

\begin{figure}[t]
\centering
\includegraphics[width=\linewidth]{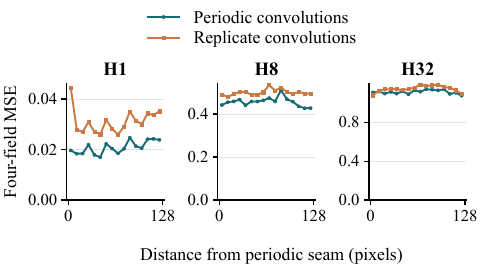}
\caption{\textbf{Error versus distance from the periodic seam.} Four-field MSE is averaged over radial positions, both sides of the seam, and test windows in eight-pixel bins. Only learned convolution padding changes. Appendix~\ref{supp:seam} details the construction.}
\label{fig:periodic_boundary_analysis}
\end{figure}

\subsection{Super-resolution and field completion}
The same model achieves the lowest MSE in all three reconstruction settings (Table~\ref{tab:restoration_main}).
SR MSE is 0.00813, slightly below BasicVSR++. $\mathrm{FC_s}$ and $\mathrm{FC_c}$ errors are 0.04147 and 0.02042, reducing RAM's errors by 48.86\% and 60.01\%, respectively.
These results support shared four-field recovery from coarse grids, sparse locations, and partial channels.
Figure~\ref{fig:restoration_qualitative} compares the same vorticity target and local region: UniPlasmaTurb recovers details closer to the target with smaller errors around several structures.
Appendix~\ref{supp:field_completion} gives per-field errors and further reconstructions.

\subsection{Ablations and structure supervision}
Table~\ref{tab:structure_evidence} compares the full model with three variants: no structure losses, full-SOL normalization, or replicated padding in learned convolutions.
The last variant retains periodic resampling and supervision operators.
The full model has the highest recall at all horizons. Removing structure supervision reduces recall by 7.94, 4.46, and 7.35 percentage points at H8, H16, and H32.
Figure~\ref{fig:structure_supervision_analysis} groups H16 predictions by true structure strength and occupied SOL area.
Predicted peaks are measured within every true region, including missed structures, so responses reflect amplitude and position errors.
The full model retains a higher mean response than the no-structure variant in all five strength groups.
Below 2.01\% SOL occupancy, recall is 48.91\%, versus 35.87\% without structure losses and 36.96\% without balancing.
Regional balancing improves recall by 2.02 and 2.51 percentage points in the two highest-occupancy groups, with larger gains when structures occupy less area.

\subsection{Role of historical information}
With fixed weights, four input slots contain one, two, or four distinct states, with repetitions as needed.
In Fig.~\ref{fig:history_information_analysis}(a), using four states instead of one reduces missing-field MSE from 0.08251 to 0.03011 for $W$ and from 0.28758 to 0.01073 for $\Phi_{\mathrm{eff}}$ (factors of 2.74 and 26.81).
Temporal changes thus help infer unobserved fields from $P,N$.
Panel (b) reports mean localization error over all true structures, using periodic distances and assigning 16 pixels to each miss.
Compared with one frame, four frames reduce error from 12.00 to 9.68 pixels at H8 and from 13.91 to 12.90 at H16. At H32, two and four frames give similar errors (13.48 and 13.56 pixels).
These results show that history aids cross-field completion and shorter-horizon localization. Appendix~\ref{supp:history} details input settings.

\subsection{Role of periodic convolution padding}
Periodic convolution padding reduces one-step error near the periodic seam (Fig.~\ref{fig:periodic_boundary_analysis}).
Within eight pixels, MSE is 0.01966 with periodic padding versus 0.04456 with replication.
The advantage also extends inward at H8. At H32, the gap narrows and some locations favor replication.

\section{Conclusion And Future Work}
\label{sec:conclusion}
This work builds a coupled four-field simulation system and dataset for edge turbulence and introduces UniPlasmaTurb for unified reconstruction and forecasting.
By preserving observations and balancing structure and background supervision, the framework achieves lower reconstruction errors and higher multi-step structure recall with one model.
Our future work will broaden physical and observation conditions and further assess long-term physical behavior through temporal structure continuity and particle transport.

{\small
\bibliographystyle{ieeenat_fullname}
\bibliography{cvpr2027_references}
}


\twocolumn[{\centering\Large\bfseries Appendices\par\medskip}]
\raggedbottom

\appendix
\setcounter{figure}{0}
\setcounter{table}{0}
\setcounter{equation}{0}
\renewcommand{\theHfigure}{supp.\arabic{figure}}
\renewcommand{\theHtable}{supp.\arabic{table}}
\renewcommand{\theHequation}{supp.\arabic{equation}}
\renewcommand{\thefigure}{S\arabic{figure}}
\renewcommand{\thetable}{S\arabic{table}}
\renewcommand{\theequation}{S\arabic{equation}}

\section{Turbulence Equations And Data Generation}
\label{supp:physics}
This section specifies the coupled equations and the simulation procedure behind the dataset. The numerical state is $U=(P,N,W,\Phi_{\mathrm{eff}})$, comprising pressure and density perturbations, generalized vorticity, and effective potential. The simulation equations generate the learning targets. Observation consistency and structure supervision are implemented by the learning framework described in Sec.~\ref{supp:implementation}.

\subsection{Variables, domain, and fixed backgrounds}
The local domain is $\Omega=[0.8,1.2]\times[-0.2,0.2]$, with radial coordinate $x$ and periodic coordinate $y$. The interface $x=1$ separates the confined edge from the scrape-off layer (SOL). All four fields are periodic in $y$ and satisfy zero Dirichlet conditions at the two radial boundaries. The interface changes the source terms without imposing an internal wall condition.

The background profiles are fixed across trajectories:
\begin{equation}
\begin{aligned}
n_0(x)=T_0(x)&=1.1-\tanh\!\left(\frac{x-0.98}{0.05}\right),\\
p_0(x)&=n_0(x)T_0(x),\\
B_\phi(x)&=\frac{B_0R_0}{R_0+ax}.
\end{aligned}
\label{supp:eq:background}
\end{equation}
Here $P$ and $N$ denote departures from $p_0$ and $n_0$, rather than the total pressure and density. The spatial factor $a$ converts coordinate derivatives to the scale used by the generator. The safety-factor profile is
\begin{equation}
q(x)=q_0+\frac{(q_a-q_0)\rho_q^2}{1-(1-\rho_q^2)^{\nu_q+1}},\qquad
\rho_q=\min(x,1).
\label{supp:eq:q}
\end{equation}
It sets the parallel wavenumber and connection length. The remaining background-dependent quantities are
\begin{equation}
\begin{aligned}
 C_\phi&=\frac{n_0m_iT_{\mathrm{ref}}}{eB_0^2},&
 c_s&=\sqrt{\frac{T_0T_{\mathrm{ref}}e}{m_i}},\\
 k_\parallel&=\frac{1}{R_0q},& L_C&=\pi R_0q,\qquad \tau_p=\frac{L_C}{c_s}.
\end{aligned}
\label{supp:eq:coefficients}
\end{equation}
The equations use singly charged ions ($Z_i=1$). The constants and initial-condition choices are listed in Table~\ref{supp:tab:physics}. Values in this section use the generator's numerical convention. The separate normalization used by the learning models is defined in Sec.~\ref{supp:protocol}.

\begin{table}[t]
\centering
\caption{Simulation constants and initial-condition choices. The elementary-charge value is retained as specified by the data generator. The spectral coefficients and phases vary by trajectory.}
\label{supp:tab:physics}
\small\setlength{\tabcolsep}{3pt}
\begin{tabular}{ll}
\toprule
Quantity & Value \\
\midrule
\multicolumn{2}{l}{\textbf{(a) Fixed parameters}}\\
Spatial scale $a$ & $0.45$\\
Magnetic/geometric constants $B_0,R_0$ & $2.5,\ 1.9$\\
Diffusivity parameter $\kappa$ & $0.2$\\
Reference scales $T_{\mathrm{ref}},n_{\mathrm{ref}}$ & $100,\ 10^{19}$\\
Charge constant $e$ & $1.609\times10^{-19}$\\
Ion mass $m_i$, charge number $Z_i$ & $3.2\times10^{-27},\ 1$\\
Resistivity parameter $\eta_0$ & $10^{-6}$\\
Safety-factor parameters $q_0,q_a,\nu_q$ & $1,\ 5,\ 4$\\
Periodic domain length $L_y$ & $0.4$\\
\midrule
\multicolumn{2}{l}{\textbf{(b) Initial pressure perturbation}}\\
Base amplitude $A_0$ & $0.002$\\
Amplitude factor $\alpha$ & $0.75,\ 1.25$\\
Radial center $x_c$ & $0.97,\ 0.985,\ 1.0$\\
Radial width $s_x$ & $0.008,\ 0.016$\\
Low-band modes & $4,7,10,13$\\
Middle-band modes & $18,26,34,42$\\
Broad-band modes & $4,10,18,26,34,42,50$\\
Phase realizations per group & $4$\\
\bottomrule
\end{tabular}
\end{table}

\subsection{Coupled evolution and regional source terms}
With $D=\kappa/a^2$, the four-field system is
\begin{equation}
\begin{aligned}
\partial_t P+\mathbf u\cdot\nabla P&=D\Delta P+S_P,\\
\partial_t N+\mathbf u\cdot\nabla N&=D\Delta N+S_N,\\
\partial_t W+\mathbf u\cdot\nabla W&=D\Delta W+S_W,\\
\nabla\cdot(C_\phi\nabla\Phi_{\mathrm{eff}})&=a^2W.
\end{aligned}
\label{supp:eq:system}
\end{equation}
The first three equations evolve the perturbations by advection, diffusion, and sources. The last equation couples vorticity and potential at each time. Its spatially varying coefficient $C_\phi(x)$ is retained inside the divergence.

Define $\chi_c=\mathbf1_{x\leq1}$, $\chi_s=\mathbf1_{x>1}$ and $\phi_N=\Phi_{\mathrm{eff}}-P/n_0$. The complete sources are
\begin{equation}
\begin{aligned}
S_P={}&-u_x\frac{\mathrm d p_0}{\mathrm d x}-\chi_s\frac{P}{\tau_p},\\
S_N={}&-u_x\frac{\mathrm d n_0}{\mathrm d x}-\chi_s\frac{N}{\tau_p},\\
S_W={}&-\frac{2T_{\mathrm{ref}}}{B_0R_0a}\partial_yP\\
&+\chi_c\frac{k_\parallel^2T_{\mathrm{ref}}}{e n_{\mathrm{ref}}\eta_0}
 \left(\phi_N-\frac{N}{n_0}\right)
 +\chi_s\frac{2n_0c_s}{L_C}\phi_N.
\end{aligned}
\label{supp:eq:sources}
\end{equation}
The pressure and density sources combine radial transport of the background gradients with SOL losses. The vorticity source combines pressure-gradient curvature drive with different current-related terms on the two sides of the interface.

The electric field and the drift velocity in model coordinates satisfy
\begin{equation}
E_i=-\frac{T_{\mathrm{ref}}}{a}\left(\partial_i\Phi_{\mathrm{eff}}-
\frac{\partial_iP}{n_0}\right),\qquad
\mathbf u=\frac{(E_y,-E_x)}{aB_\phi},
\label{supp:eq:velocity}
\end{equation}
where $i\in\{x,y\}$. Pressure and potential gradients therefore determine the velocity that transports the three evolving fields. Because $n_0$ varies radially, the radial electric field uses $\partial_xP/n_0$, rather than $\partial_x(P/n_0)$.

\subsection{Initial perturbations and numerical generation}
Each trajectory starts from a pressure perturbation of the form
\begin{equation}
\begin{split}
P(x,y,0)={}&A_0g\exp\!\left[-\frac{(x-x_c)^2}{2s_x^2}\right]\\
&\times\sum_{j=1}^{8}w_j\cos\!\left(\frac{2\pi m_jy}{L_y}+\varphi_j\right).
\end{split}
\label{supp:eq:initial}
\end{equation}
Initially $N=W=\Phi_{\mathrm{eff}}=0$. The generator reserves eight mode slots. For a spectral family with $m$ active modes, each active weight is $1/\sqrt{m}$ and all other weights are zero. The gain for an envelope centered at $x_c$ with width $s_x$ is
\begin{equation*}
  g=\alpha\sqrt{\frac{I_{\mathrm{ref}}}{I(x_c,s_x)}},\quad
  I(x_c,s_x)=\int_{0.8}^{1.2}e^{-(x-x_c)^2/s_x^2}\,\mathrm{d}x .
\end{equation*}
The reference integral $I_{\mathrm{ref}}$ uses $x_c=0.975526315789$ and $s_x=0.0107368421053$. At fixed $\alpha$, the initial pressure perturbations therefore have the same squared spatial norm. The three centers, two widths, two amplitude factors, and three spectral families in Table~\ref{supp:tab:physics} define 36 groups. Each group contains four phase realizations, all assigned to the same split.

Spatial discretization uses quadratic Lagrange elements on a fixed mesh of 24,912 triangles with 12,657 vertices and fourth-order weak-form quadrature. A fully coupled direct solver advances the system with the backward differentiation formula (BDF), using adaptive steps of order at most two. The prescribed initial step is $1/6000\,\mu\mathrm{s}$ and the maximum internal step is $1/6\,\mu\mathrm{s}$.

The solution is sampled every $1/6\,\mu\mathrm{s}$, giving 1,201 snapshots per trajectory over $0$--$200\,\mu\mathrm{s}$. Each snapshot is interpolated onto a regular $257\times257$ grid. Removing the duplicate periodic endpoint leaves $n_y\times n_x=256\times257$ values per field, with $\Delta x=\Delta y=0.4/256$ and both radial endpoints retained.

\section{Dataset, Observations, And Evaluation Metrics}
\label{supp:protocol}
The tasks are super-resolution (SR), field completion (FC), and forecasting (FP). FC comprises sparse spatial completion (FC$_s$) and missing-field completion (FC$_c$).

\subsection{Splits and normalization}
The group-level partition contains 96 training, 24 validation, and 24 test trajectories, corresponding to 24, six, and six initial-condition groups. Weight training uses one predetermined phase realization from each training group, giving 24 training trajectories. Input means $\mu_{24,c}$ and standard deviations $\sigma_{24,c}$ are fitted on their 601 stride-two snapshots and frozen throughout training and evaluation.

Reported errors use common channelwise means $\mu_{96,c}$ and standard deviations $\sigma_{96,c}$ from the complete 96-trajectory training split. A prediction is first converted back through its model normalization and then to these reporting units. The conversion preserves the physical prediction while putting all methods on the same scale. No field values are clipped. The full training split also determines the structure threshold and the analysis bins in Sec.~\ref{supp:mechanisms}.

\begin{table}[t]
\centering
\caption{Data and evaluation protocol. One forecasting step spans two native snapshot intervals. All reported test comparisons use the same 72 windows.}
\label{supp:tab:protocol}
\small
\begin{tabular}{lp{4.25cm}}
\toprule
Item & Setting\\
\midrule
Trajectory split & 96 training / 24 validation / 24 test\\
Weight-training subset & 24 trajectories, one phase per training group\\
Input statistics & 24 training trajectories, 601 stride-two frames\\
Reporting statistics & Complete 96-trajectory training split\\
Field grid & $256\times257$, four channels\\
Native / model step & $1/6$ / $1/3\,\mu\mathrm{s}$\\
Test anchors $a$ & $232,520,920$ per trajectory\\
History indices & $a-8,a-6,a-4,a-2$\\
Reconstruction target & Native index $a-2$\\
FP target at step $h$ & Native index $a+2(h-1)$\\
SR observation & Four fields at $128\times129$\\
FC$_s$ observation & 6,579 locations, shared over fields and history\\
FC$_c$ observation & Complete $P,N$ fields\\
FC$_c$ target & Missing $W,\Phi_{\mathrm{eff}}$ fields\\
\bottomrule
\end{tabular}
\end{table}

\subsection{Time windows and observation operators}
\begin{figure*}[t]
\centering
\includegraphics[width=\textwidth]{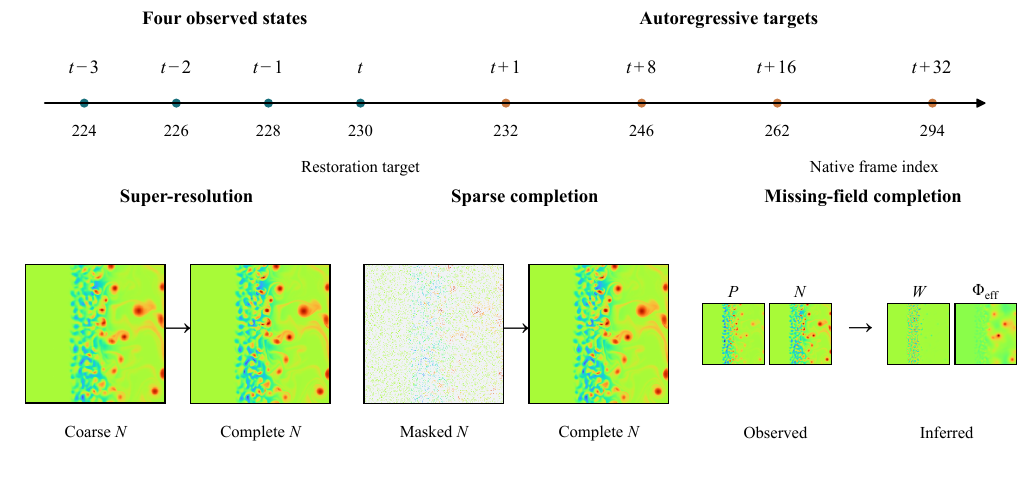}
\caption{\textbf{Time windows and observation types.} The timeline shows the native indices for anchor $a=232$, with schematic spacing. Reconstruction targets the last observed state, while forecasting advances beyond it. The lower panels illustrate coarse-grid, sparse-position, and missing-field observations using test trajectory G08/P1. Each task uses the same indexing for its observed history.}
\label{supp:fig:protocol}
\end{figure*}

Let $a$ denote the zero-based native index of the first future frame. UniPlasmaTurb receives history indices $(a-8,a-6,a-4,a-2)$. Reconstruction targets $a-2$, while prediction step $h$ targets $a+2(h-1)$. Thus H32 is $32/3\,\mu\mathrm{s}$ after the last observation. The fixed test anchors $a\in\{232,520,920\}$ produce 72 windows over 24 test trajectories. Figure~\ref{supp:fig:protocol} illustrates the temporal indexing and observation types. Baseline-specific history lengths are given in Sec.~\ref{supp:baselines}.

For SR, each history frame is filtered with the separable kernel $(1,4,6,4,1)/16$ and downsampled by two. Filtering wraps in $y$ and replicates the radial edges in $x$. The retained radial endpoints use the periodically filtered endpoint values without radial smoothing. This gives $128\times129$ observations and defines the same operator $R$ used for output consistency.

For sparse completion (FC$_s$), 6,579 of the 65,792 spatial positions are sampled without replacement. The same mask applies to all four fields and history frames within a window. The mask is generated from a fixed trajectory/anchor seed and reused across methods. For missing-field completion (FC$_c$), $P$ and $N$ are observed everywhere, and $W$ and $\Phi_{\mathrm{eff}}$ are hidden. Missing entries are zeroed before model input and marked by availability masks. Forecasting starts from complete observed states and receives no future measurements.

\subsection{Field errors and structure matching}
\label{supp:matching}
For reporting, define $Z_c=(U_c-\mu_{96,c})/\sigma_{96,c}$. The mean squared error (MSE) over the four fields in one window is
\begin{equation}
\mathrm{MSE}=\frac{1}{4n_yn_x}\sum_{c,y,x}(\widehat Z_c-Z_c)^2.
\end{equation}
A per-field error retains only the corresponding channel. FC errors average only over missing channel--pixel entries, so FC$_c$ evaluates $W$ and $\Phi_{\mathrm{eff}}$. Errors are averaged over windows within each trajectory and then over trajectories. With three windows per trajectory, this equals the mean over all 72 windows.

Density fluctuations are computed before restricting to the SOL:
\begin{equation}
N'_Z(y,x)=\frac{N(y,x)-n_y^{-1}\sum_{y'}N(y',x)}{\sigma_{96,N}}.
\end{equation}
The SOL comprises radial columns $129,\ldots,256$, corresponding to $x>1$. The threshold $\tau=0.7661558063$ is the 95th percentile of $|N'_Z|$ over all 96 training trajectories at native frames $120,240,\ldots,1200$.

A positive structure is an eight-connected component of $N'_Z\geq\tau$ containing at least 36 pixels. Connectivity wraps in $y$. The object's location is its maximum-amplitude pixel. For a predicted peak and a true peak, the distance in pixels is
\begin{equation}
d=\sqrt{(x_p-x_t)^2+\min(|y_p-y_t|,n_y-|y_p-y_t|)^2}.
\label{supp:eq:distance}
\end{equation}
Pairs within 16 pixels are sorted by distance and greedily matched one-to-one. Detection is evaluated independently at each forecast endpoint. With accumulated true, predicted, and matched counts $n_T,n_P,n_M$, recall and the false discovery rate (FDR) are
\begin{equation}
R=\frac{n_M}{n_T},\qquad \mathrm{FDR}=\frac{n_P-n_M}{n_P}.
\end{equation}
Counts are pooled before taking ratios. A zero denominator gives an undefined ratio. FDR measures unmatched predicted structures and complements recall's measure of recovered true structures.

For the localization analysis, every true structure contributes to the denominator. Matched structures contribute their distance in Eq.~\eqref{supp:eq:distance}, and each missed structure contributes 16 pixels:
\begin{equation}
E_{\mathrm{loc}}=\frac{\sum_{j\in\mathcal M}d_j+16(n_T-n_M)}{n_T}.
\label{supp:eq:localization}
\end{equation}
This metric reflects both detection failures and localization errors. It is distinct from the mean distance over matched structures alone.

\section{Implementation, Training, And Baseline Adaptations}
\label{supp:implementation}
\subsection{Shared network and input representation}
\label{supp:architecture}
The unified input has 41 channels: 16 historical field channels, 16 availability channels, five geometry channels, and a four-way task code. The geometry channels encode radial position, sine and cosine of periodic position, and two grid-scale constants. For a target grid of size $n_y\times n_x$, the latter are $\log_2(128/n_y)$ and $\log_2(128/(n_x-1))$. The task code is broadcast over the spatial grid.

SR observations are lifted by bilinear interpolation on a periodically extended grid. Their availability channels are one on the prepared grid, and the original coarse observations are retained for projection. FC uses zero-filled fields and the observation mask. FP uses complete historical states. These inputs share one convolutional adapter, residual U-Net, and four-channel output head.

The encoder/decoder widths are $64,128,256,128,64$, with two residual blocks per stage. Each block has two group-normalization, GELU, and $3\times3$ convolution operations. Learned convolutions wrap in $y$ and replicate radial boundary values. Low-pass restriction preserves radial endpoints during downsampling. Periodic bilinear interpolation connects the two ends of $y$ during upsampling. Decoder features concatenate with encoder features at the same scale, followed by an additional highest-resolution encoder--decoder addition. A zero-initialized $1\times1$ head predicts an increment relative to the latest prepared frame.

\begin{table}[t]
\centering
\caption{UniPlasmaTurb architecture and training settings. The same 4,787,652 parameters serve all four task settings.}
\label{supp:tab:implementation}
\small\setlength{\tabcolsep}{3pt}
\begin{tabular}{ll}
\toprule
\multicolumn{2}{l}{\textbf{(a) Architecture}}\\
Input adapter & $41\rightarrow64$, $3\times3$\\
Encoder grids & $256\times257$, $128\times129$\\
Bottleneck grid & $64\times65$\\
Stage widths & $64,128,256,128,64$\\
Residual blocks per stage & $2$\\
Normalization / activation & Group normalization / GELU\\
Output head & $64\rightarrow4$, $1\times1$\\
Learned convolution padding & Periodic $y$, replicated $x$\\
\midrule
\multicolumn{2}{l}{\textbf{(b) Joint training}}\\
Optimizer / weight decay & AdamW / $0.01$\\
Effective batch / epochs & $128$ / $32$\\
Updates per epoch / total & $111$ / $3,552$\\
Learning-rate warmup & $10^{-4}\rightarrow10^{-3}$, 5 epochs\\
Subsequent schedule & Cosine decay toward $10^{-4}$\\
Task weights: SR, FC$_s$, FC$_c$, FP & $1/3,1/6,1/6,1/3$\\
Data / amplitude / occupancy & $1.1,0.1,0.05$\\
Occupancy neighborhood & $7\times7$\\
Training threshold / temperature & $0.762674093$ / $0.149318341$\\
\bottomrule
\end{tabular}
\end{table}

\subsection{Observation-consistent output projection}
\label{supp:projection}
Let $R_y$ and $R_x$ denote the one-dimensional filtering and decimation matrices, including the periodic and endpoint rules in Sec.~\ref{supp:protocol}. Their separable action defines $R$. The fixed right inverse is
\begin{equation}
R^\dagger=R^{\mathsf T}(RR^{\mathsf T})^{-1}.
\end{equation}
The implementation solves the one-dimensional Gram systems without ridge regularization and applies the resulting separable operators. Since $R$ has full row rank, $RR^\dagger=I$. For raw output $Z$ and coarse observation $Y$, the corrected output is
\begin{equation}
\widehat U=Z+R^\dagger(Y-RZ).
\end{equation}
Applying $R$ gives $R\widehat U=RZ+Y-RZ=Y$ up to numerical precision. For FC, $\widehat U=M\odot Y+(1-M)\odot Z$ directly retains observed entries. FP uses $Z$ because the future state has no available measurements. The same output rules are applied during training and inference.

\begin{table*}[!t]
\centering
\caption{Baseline configurations used in the main comparisons. Frames count distinct observed states used by the model. Training updates are optimizer updates. Time-limited stages retain their final checkpoints.}
\label{supp:tab:baselines}
\small\setlength{\tabcolsep}{4pt}
\begin{tabular}{llcp{2.6cm}p{7.1cm}}
\toprule
Method & Task & Frames & Training endpoint\newline (optimizer updates) & Input or inference adaptation\\
\midrule
FNO~\cite{li2021fourier} & FP & 4 & 128 epochs, 14,208 & 16 historical channels to four next-state channels\\
CNO~\cite{raonic2023cno} & FP & 4 & 64 epochs, 7,104 & Rectangular padded grid, antialiasing operators\\
PDE-Transformer~\cite{holzschuh2025pdetransformer} & FP & 4 & 128 epochs, 14,208 & Four-field next-state prediction with autoregressive feedback\\
LinearNO~\cite{hu2025linearno} & FP & 4 & 32 epochs, 3,552 & One-step MSE, four-frame adaptation\\
CFO-U-Net~\cite{hou2026cfo} & FP & 1 & 128 epochs, 7,040 & Continuous-time velocity model, RK4 with two substeps\\
DLL-NO~\cite{park2026dll} & FP & 1 & 28,447 / 23,995 & Operator and diffusion stages, mean of 32 sampled trajectories\\
\midrule
BasicVSR++~\cite{chan2022basicvsrpp} & SR & 4 & 32 epochs, 3,552 & Low-resolution history to final high-resolution frame\\
HiNOTE~\cite{luo2024hinote} & SR & 4 & 32 epochs, 3,552 & 16-to-four channel adaptation, L1 objective\\
UCAN~\cite{tan2026ucan} & SR & 1 & 14,208 & Four physical channels, scale-two reconstruction\\
IRNO~\cite{liu2026irno} & SR & 4 & 28,416 / 133,200 & Trained FNO base followed by six-step residual refinement\\
RAM~\cite{supp_terris2025ram} & SR/FC & 4 & 32 epochs, 3,552 & Joint restoration, native conditioning operators\\
4DVarNet~\cite{beauchamp2023varnet} & FC & 4 & 32 epochs, 3,552 & Ten learned optimization steps, final-state output\\
FLRONet~\cite{dang2025flronet} & FC$_s$ & 4 & 32 epochs, 3,552 & Voronoi sensor embedding and final-time query\\
LANO~\cite{hou2026lano} & FC$_s$ & 4 & 32 epochs, 3,552 & Masked historical input, latent autoregressive modeling\\
PhySense~\cite{ma2025physense} & FC$_s$ & 1 & 9,119 & Official $2\times2$ patches, ten sampling steps\\
OmniField~\cite{valencia2026omnifield} & FC$_c$ & 1 & 100,000 & Official FP32 core, single-frame $P,N\rightarrow W,\Phi_{\mathrm{eff}}$\\
\bottomrule
\end{tabular}
\end{table*}

\subsection{Loss implementation and optimization}
Data losses are computed on the final output. SR and FP use complete-field MSE, and FC uses the mean over missing entries. The structure terms operate on density fluctuations over the full SOL. Let $p,n$ be predicted and true fluctuations, $A=\{i:n_i>\tau\}$, and
\begin{equation}
\mathcal B_B(e)=\frac{\sum_{i\in B}e_i}{\max(|B|,1)}.
\end{equation}
The sum over an empty region is zero. Each region mean is computed per sample before averaging over the batch. The exceedance (amplitude) loss is
\begin{equation}
\mathcal L_{\mathrm{exc}}=\tfrac12\bigl[\mathcal B_A((p-n)^2)+
\mathcal B_{A^c}([p-\tau]_+^2)\bigr].
\end{equation}
For occupancy, $o(v)=\mathcal A_7[\sigma((v-\tau)/T)]$ uses sigmoid responses followed by a $7\times7$ local mean. Pooling wraps in $y$ and replicates the radial SOL edges. With $S=\{i:(\mathcal A_7\mathbf1_A)_i>0\}$,
\begin{equation}
\mathcal L_{\mathrm{occ}}=\tfrac12\bigl[\mathcal B_S((o(p)-o(n))^2)+
\mathcal B_{S^c}((o(p)-o(n))^2)\bigr].
\end{equation}
The reporting-unit threshold and temperature are $0.7661558063$ and $0.15$. Conversion to the input normalization gives the values in Table~\ref{supp:tab:implementation}. Separate region means prevent the larger background from dominating solely through pixel count. For FC$_c$, density is observed and copied exactly, so both density-structure losses are zero.

Every optimizer update constructs the four task settings from the same batch of histories, accumulates their weighted gradients, and updates the shared parameters. SR and FC target the last observed state, while FP targets the next state. There are 14,328 candidate one-step windows per epoch, indexed by $a=8,10,\ldots,1200$ over the 24 training trajectories. Shuffling and dropping the incomplete final batch retain 14,208 windows per epoch, giving 454,656 forecasting-window presentations over 32 epochs. The main model uses no gradient clipping or mixed-precision autocast. Training permits TF32, while test evaluation uses FP32 with TF32 disabled.

Experiments run on NVIDIA A800-SXM4-80GB GPUs with dual Intel Xeon Gold 6348 processors. The main model's 30.78 GPU-hours cover joint training for all tasks through epoch 32. Baseline costs cover their corresponding trained components and stages. These are training costs, rather than inference latency, and budgets and task coverage differ across methods.

\subsection{Baseline configurations and task adaptations}
\label{supp:baselines}
Table~\ref{supp:tab:baselines} identifies the baseline configurations used in Sec.~\ref{sec:experiments}. Every method is trained on the plasma data with its stated observation interface.

\paragraph{Forecasting.}
FNO, CNO, PDE-Transformer, and LinearNO stack the four historical states and predict the next four-field state. Predictions are fed back for later steps. CNO pads the grid to $256\times272$ internally, uses rectangular antialiasing operations, and disables batch normalization. LinearNO uses one-step MSE with the adapted warmup/cosine schedule.

CFO-U-Net receives the latest state and time. It fits a velocity field to quintic interpolation of training trajectories and advances with fourth-order Runge--Kutta (RK4) integration using two substeps per model interval. Absolute time is normalized over the 601-frame sequence, with corresponding velocity rescaling. DLL-NO uses its native single-state interface. Its operator and diffusion stages contain 28,447 and 23,995 updates, respectively, under the established time-budget run. Reported predictions are the mean field of 32 sampled trajectories.

\paragraph{Restoration.}
BasicVSR++ reconstructs the last high-resolution state from four low-resolution observations. HiNOTE uses its L1 objective without the optional frequency-prior loss. UCAN receives the latest four-channel low-resolution state. IRNO first trains an SR-compatible FNO base and then freezes it while learning the residual refiner. Six refinement steps are used at inference. RAM jointly trains SR and both FC settings with its conditioning components. It uses width 16 and 28 backbone blocks, without an additional hard output projection. The 4DVarNet adaptation returns its tenth learned solver state. FLRONet uses deterministic Voronoi embedding of sparse sensors, while LANO receives the four masked states. PhySense uses the latest observed state, fixed sparse masks, and its native $2\times2$ patch construction, with a one-column pad to $256\times258$.

OmniField follows the official core implementation in FP32 and is trained from random initialization for 100,000 updates. The plasma interface supplies one complete $P,N$ frame to infer $W,\Phi_{\mathrm{eff}}$. Training uses batch size eight and AdamW with weight decay $10^{-4}$ and gradient clipping at norm one. The learning rate warms from $8\times10^{-6}$ to $8\times10^{-5}$ over 1,000 updates, then decays by cosine to $8\times10^{-6}$. The registered model has 41,775,492 parameters and takes 30.12 GPU-hours.

\section{Complete Comparative Results And Qualitative Examples}
\label{supp:comparisons}
\subsection{Multi-step forecasting by physical field}
\label{supp:field_forecast}
At H16, UniPlasmaTurb recovers $47.39\%$ of the true positive density structures. DLL-NO recovers $0.65\%$ despite lower MSE in each of the four fields (Tables~\ref{tab:fp_main} and~\ref{supp:tab:forecast}). Field error and structure recall thus rank these forecasts differently. The false discovery rate (FDR) gives the fraction of predicted positive density structures left unmatched.

\begin{table*}[!t]
\centering\small
\caption{Per-field forecasting MSE and positive density structure FDR (\%) on the 72 test windows. Lower is better in every column. Each method uses the same checkpoint as in Table~\ref{tab:fp_main}.}
\label{supp:tab:forecast}
\setlength{\tabcolsep}{6pt}
\begin{tabular}{llrrrrr}
\toprule
Method & Horizon & $P$ & $N$ & $W$ & $\Phi_{\mathrm{eff}}$ & FDR \\
\midrule
FNO & H8 & 0.36190 & 0.34190 & 1.69464 & 0.67875 & 53.53 \\
 & H16 & 0.63665 & 0.56619 & 1.61105 & 1.18523 & 69.29 \\
 & H32 & 0.98792 & 0.88483 & 1.52385 & 1.45518 & 74.21 \\
\midrule
CNO & H8 & 0.27317 & 0.27531 & 1.48754 & 0.40099 & 44.01 \\
 & H16 & 0.66003 & 0.62361 & 2.21532 & 0.95298 & 63.22 \\
 & H32 & 1.46316 & 1.29054 & 2.51511 & 3.06670 & 74.48 \\
\midrule
PDE-Transformer & H8 & 0.20472 & 0.21583 & 1.31296 & 0.34458 & 33.92 \\
 & H16 & 0.36952 & 0.35095 & 1.96749 & 0.57264 & 55.96 \\
 & H32 & 0.53216 & 0.44006 & 2.42417 & 0.69054 & 63.95 \\
\midrule
LinearNO & H8 & 0.32307 & 0.27947 & 1.63476 & 0.43946 & 59.48 \\
 & H16 & 0.47697 & 0.37829 & 1.65080 & 0.52405 & 78.10 \\
 & H32 & 0.62089 & 0.49408 & 1.71495 & 0.57017 & 80.10 \\
\midrule
CFO-U-Net & H8 & 0.34102 & 0.35125 & 2.05142 & 0.69071 & 63.82 \\
 & H16 & 0.43561 & 0.41054 & 2.27949 & 0.77656 & 61.90 \\
 & H32 & 0.49810 & 0.44386 & 2.31988 & 0.79562 & 63.86 \\
\midrule
DLL-NO & H8 & 0.19216 & 0.20038 & 1.11686 & 0.33907 & 24.69 \\
 & H16 & 0.25393 & 0.26171 & 1.22104 & 0.48798 & 45.45 \\
 & H32 & $6.107\times10^{3}$ & $2.331\times10^{3}$ & $4.322\times10^{3}$ & $1.494\times10^{4}$ & 91.80 \\
\midrule
UniPlasmaTurb & H8 & 0.16539 & 0.19656 & 1.18084 & 0.28350 & 28.27 \\
 & H16 & 0.33656 & 0.35311 & 2.01245 & 0.62169 & 54.54 \\
 & H32 & 0.50369 & 0.46691 & 2.67473 & 0.79595 & 63.21 \\
\bottomrule
\end{tabular}
\end{table*}

At H32, DLL-NO's four-field MSE is below one in 68 of the 72 test windows. The four remaining errors, approximately $14.85$, $3366.77$, $73872.09$, and $421234.89$, dominate the reported mean.

\begin{figure*}[t]
\centering
\includegraphics[width=\textwidth]{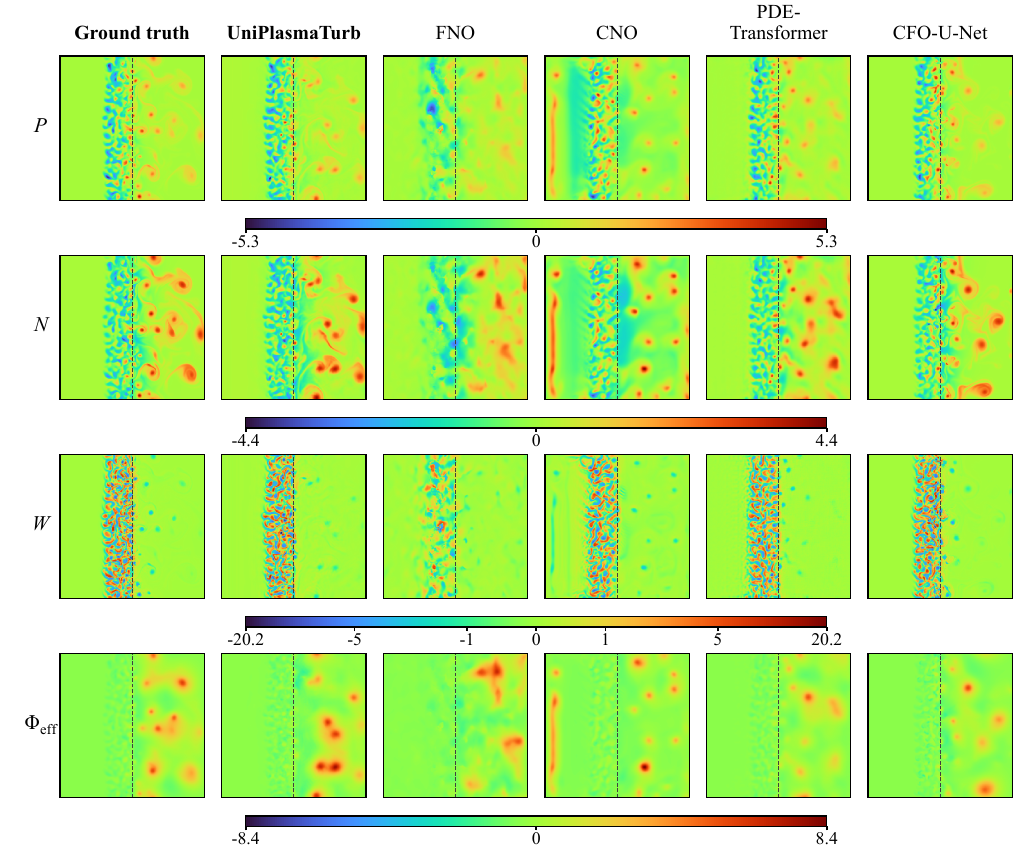}
\caption{\textbf{Four-field forecasts at H16.} Pressure, density, vorticity, and effective potential for test trajectory G08/P4 at anchor 232. Each row shares a color scale spanning all displayed values. Vorticity uses $\operatorname{asinh}(v)$ for normalized values $v$, while the other fields use linear scales. Ticks retain the original normalized units. Dashed lines mark the SOL interface. All predictions use the same checkpoints as in Fig.~\ref{fig:forecast_qualitative}.}
\label{supp:fig:forecast}
\end{figure*}

Figure~\ref{supp:fig:forecast} places the H16 density forecast from Fig.~\ref{fig:forecast_qualitative} alongside the pressure, vorticity, and effective-potential forecasts for the same window. The vorticity row uses an inverse hyperbolic sine scale to make weak fluctuations visible across its wider range.

\subsection{Super-resolution and field completion}
\label{supp:field_completion}
UniPlasmaTurb achieves the lowest MSE on each hidden field in Table~\ref{supp:tab:restoration}. Given only pressure and density, it reconstructs vorticity with an MSE of $0.03011$ and effective potential with an MSE of $0.01073$. With sparse observations, it has the lowest error at unobserved positions for all four fields. In SR, UniPlasmaTurb has the lowest four-field mean MSE, while BasicVSR++ has the lowest density MSE.

\begin{table*}[!t]
\centering\small
\caption{Per-field reconstruction MSE. SR uses complete fields, FC$_s$ uses unobserved spatial positions, and FC$_c$ uses the two missing channels. Observed channels are marked by dashes. All entries use the same 72 test windows.}
\label{supp:tab:restoration}
\setlength{\tabcolsep}{6pt}
\begin{tabular}{llrrrr}
\toprule
Task & Method & $P$ & $N$ & $W$ & $\Phi_{\mathrm{eff}}$ \\
\midrule
SR & BasicVSR++ & 0.00093 & 0.00102 & 0.03083 & 0.00005 \\
 & HiNOTE & 0.00233 & 0.00279 & 0.06776 & 0.00010 \\
 & UCAN & 0.00095 & 0.00106 & 0.03172 & 0.00007 \\
 & IRNO & 0.00117 & 0.00124 & 0.03067 & 0.00012 \\
 & RAM & 0.00542 & 0.00640 & 0.06591 & 0.00157 \\
 & UniPlasmaTurb & 0.00093 & 0.00104 & 0.03051 & 0.00002 \\
\midrule
FC$_s$ & RAM & 0.01932 & 0.02083 & 0.28072 & 0.00348 \\
 & 4DVarNet & 0.02091 & 0.02293 & 0.27197 & 0.00864 \\
 & FLRONet & 0.07225 & 0.05389 & 0.72695 & 0.00618 \\
 & LANO & 0.03600 & 0.03194 & 0.51270 & 0.00707 \\
 & PhySense & 0.37723 & 0.28958 & 1.93071 & 0.14133 \\
 & UniPlasmaTurb & 0.00897 & 0.01044 & 0.14545 & 0.00102 \\
\midrule
FC$_c$ & RAM & --- & --- & 0.07204 & 0.03006 \\
 & 4DVarNet & --- & --- & 0.06590 & 0.06599 \\
 & OmniField & --- & --- & 1.69405 & 0.25648 \\
 & UniPlasmaTurb & --- & --- & 0.03011 & 0.01073 \\
\bottomrule
\end{tabular}
\end{table*}

\begin{figure*}[!tp]
\centering
\includegraphics[width=0.97\textwidth]{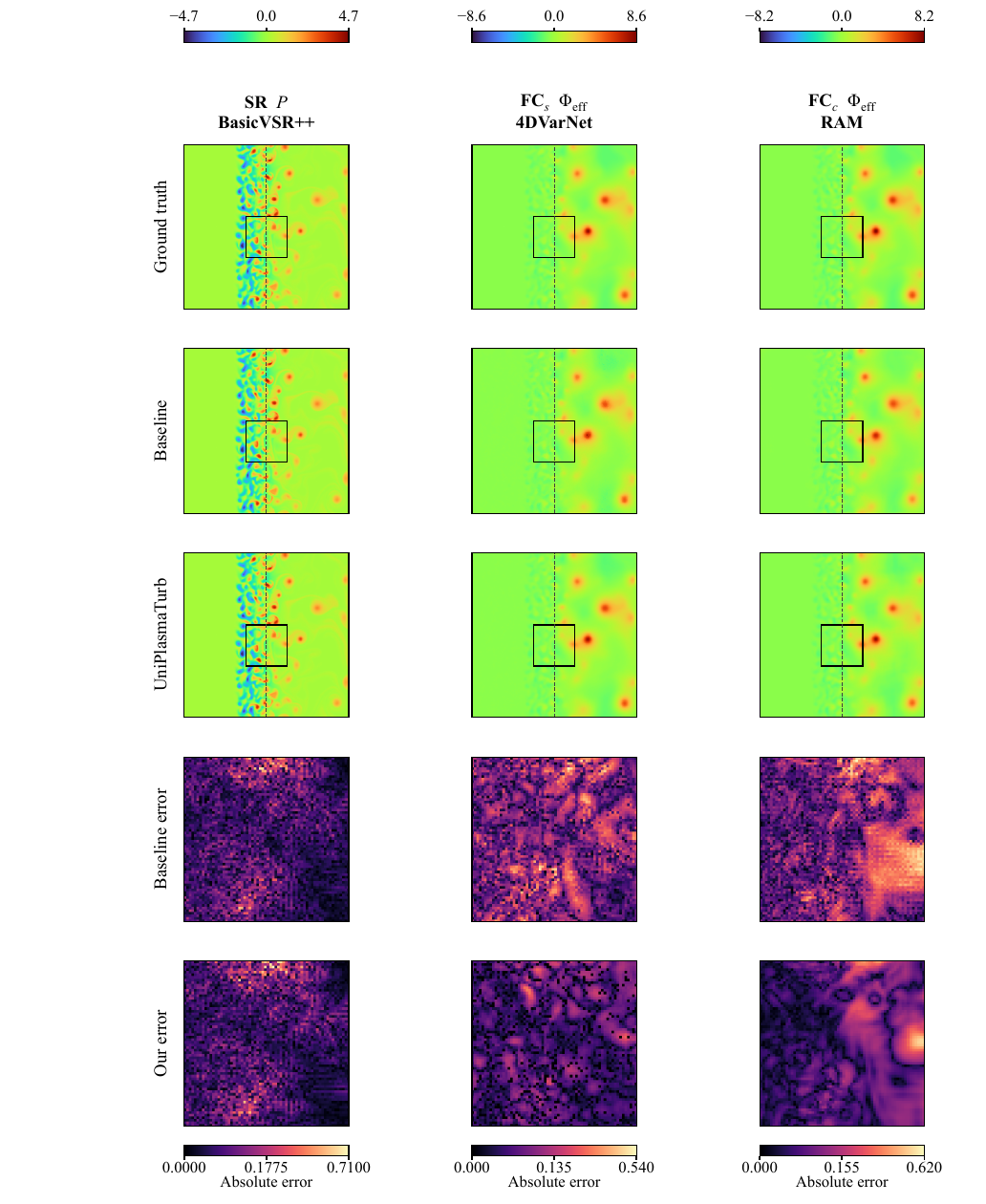}
\caption{\textbf{Additional reconstruction examples.} Test trajectory G08/P1 at anchor 232, targeting native frame 230. Columns show SR pressure, FC$_s$ effective potential, and FC$_c$ effective potential. Each column uses a shared field scale. Boxes identify the common region enlarged in the two error rows. Absolute-error scales use a square-root color mapping, with ticks showing actual errors.}
\label{supp:fig:restoration}
\end{figure*}

Figure~\ref{supp:fig:restoration} adds pressure and effective-potential reconstructions for the G08/P1 window shown in Fig.~\ref{fig:restoration_qualitative}. The error maps show every pixel in the enlarged region. For sparse completion, the table averages error only at unobserved positions.

\twocolumn[{
\section{Ablations And Mechanism Analyses}
\label{supp:mechanisms}
\subsection{Structure recall and reconstruction at epoch 32}
\label{supp:ablation_settings}
\begin{minipage}{\textwidth}
\begin{suppanchoredtable}
\caption{Epoch-32 ablations on the 72 test windows. FP cells report MSE / recall (\%) / FDR (\%). Reconstruction MSE uses complete fields for SR and hidden entries for FC.}
\label{supp:tab:ablation}
\setlength{\tabcolsep}{6pt}
\begin{tabular}{lccc}
\toprule
\multicolumn{4}{l}{\textbf{(a) Multi-step forecasting: MSE$\downarrow$ / recall$\uparrow$ / FDR$\downarrow$}} \\
Variant & H8 & H16 & H32 \\
\midrule
Full model & 0.4566 / 73.99 / 28.27 & 0.8310 / 47.39 / 54.54 & 1.1103 / 38.70 / 63.21 \\
No structure supervision & 0.4560 / 66.05 / 30.63 & 0.7682 / 42.93 / 53.03 & 0.9804 / 31.35 / 59.10 \\
No region balancing & 0.4458 / 68.12 / 28.13 & 0.7609 / 43.26 / 53.34 & 0.9428 / 36.86 / 62.03 \\
Replicated convolution padding & 0.5006 / 65.94 / 33.41 & 0.8688 / 39.35 / 60.61 & 1.1435 / 34.27 / 68.20 \\
\midrule\multicolumn{4}{l}{\textbf{(b) Reconstruction MSE$\downarrow$}} \\
Variant & SR & FC$_s$ & FC$_c$ \\
\midrule
Full model & 0.00813 & 0.04147 & 0.02042 \\
No structure supervision & 0.00978 & 0.05919 & 0.02735 \\
No region balancing & 0.00907 & 0.05190 & 0.02394 \\
Replicated convolution padding & 0.01019 & 0.06313 & 0.03085 \\
\bottomrule
\end{tabular}
\end{suppanchoredtable}
\end{minipage}
\vspace{1em}
}]
The full model has the highest positive density structure recall at H8, H16, and H32 and the lowest reconstruction MSE for SR, FC$_s$, and FC$_c$ among the four variants in Table~\ref{supp:tab:ablation}. At H8, the full model recovers $73.99\%$ of true structures, compared with $66.05\%$ without structure supervision and $68.12\%$ without region balancing.

The no-structure variant removes both the amplitude and local occupancy losses. The no-balancing variant retains both pixelwise penalties but averages each over the entire SOL. This weights the two regions by area and changes the gradient scale. The padding variant uses replicated $y$ boundaries in the 25 learned convolution layers. Radial padding, resampling, and local occupancy pooling follow the full model.

All variants were evaluated at epoch 32. The full model used its original learning-rate schedule. After training-loss spikes, the three ablations resumed from stable checkpoints with the remaining learning rates scaled by one quarter. The no-structure variant resumed from epoch three without gradient clipping. The no-balancing and replicated-padding variants resumed from epochs 14 and seven, respectively, with gradient-norm clipping at one. Neither of these two runs triggered clipping during recovery.

\newpage
At H32, removing the structure losses lowers four-field MSE from $1.1103$ to $0.9804$ and FDR from $63.21\%$ to $59.10\%$, while recall falls from $38.70\%$ to $31.35\%$. The lower mean field error therefore accompanies fewer recovered density structures. The H16 analysis below examines where the full model gains recall.

\begin{figure*}[p]
\centering
\includegraphics[width=\textwidth]{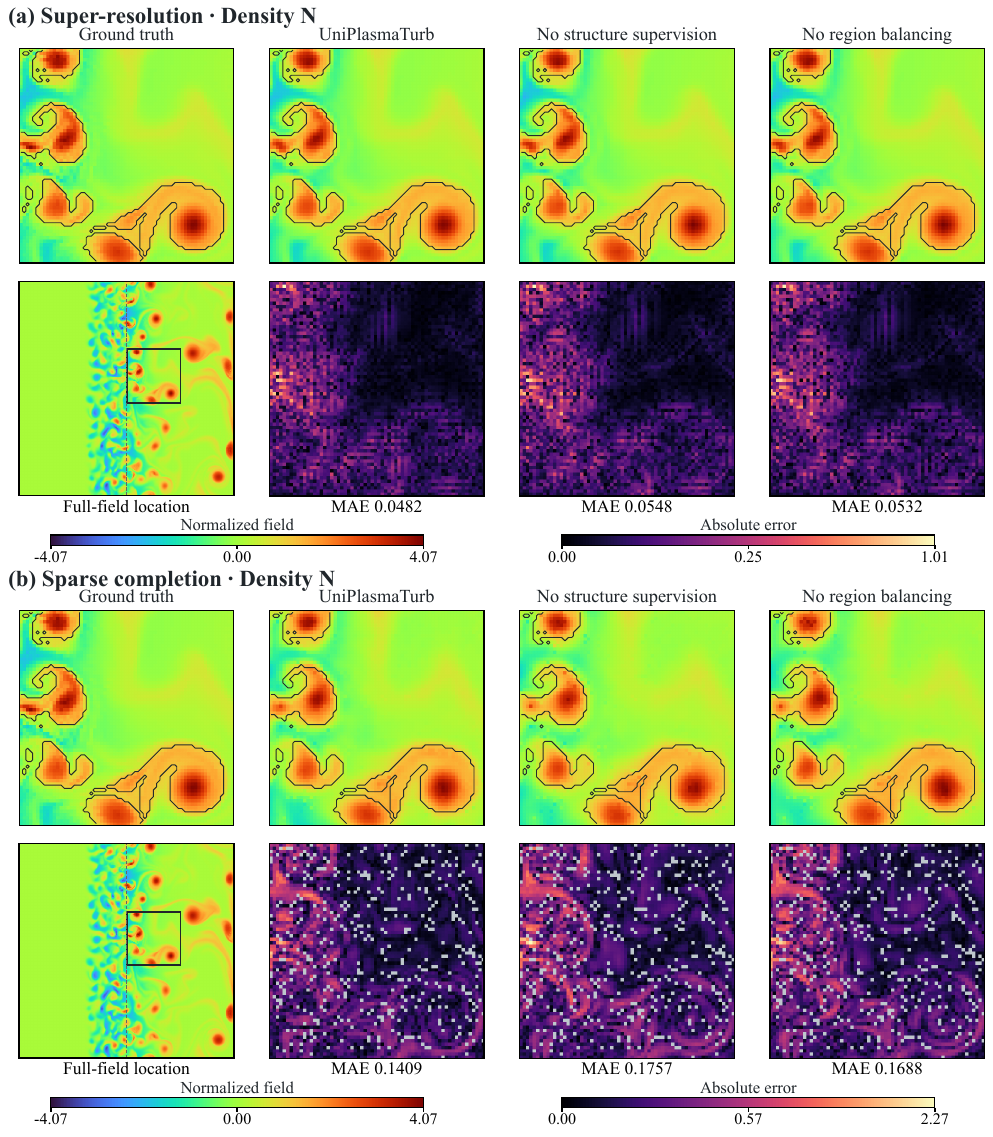}
\caption{\textbf{Density reconstruction with structure-supervision ablations.} Test trajectory G08/P1, anchor 232. Each task shows a common $64\times64$ crop, its location in the full field, and absolute errors. The crop maximizes true positive-fluctuation energy within the SOL. Black contours mark true above-threshold fluctuation regions. Fields and errors share scales across variants. Error colors use a square-root mapping with ticks in normalized units. Gray FC$_s$ pixels are observed and excluded from the reported crop MAE.}
\label{supp:fig:density_ablation}
\end{figure*}

\begin{figure*}[p]
\centering
\includegraphics[width=\textwidth]{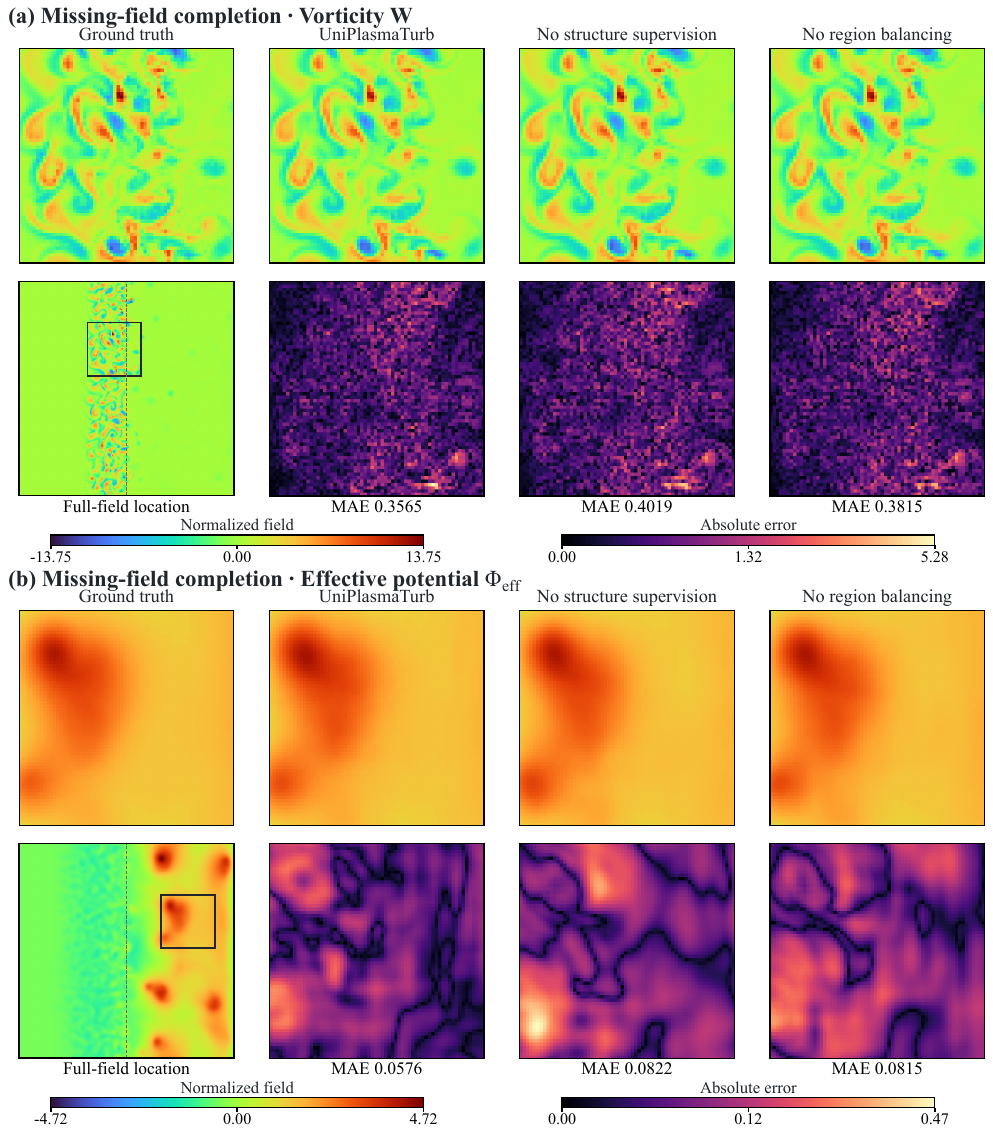}
\caption{\textbf{Missing-field reconstruction with structure-supervision ablations.} Test trajectory G08/P1, with anchor 232 in (a) and 920 in (b). Observed $P,N$ condition recovery of $W,\Phi_{\mathrm{eff}}$. For each field, the common $64\times64$ crop maximizes the true field's local squared amplitude. Lower-left maps locate the crops. Field scales are shared across variants, and error colors use a shared square-root mapping with ticks in normalized units. MAE is computed within each crop.}
\label{supp:fig:missing_ablation}
\end{figure*}

\paragraph{Reconstruction examples.}
Figures~\ref{supp:fig:density_ablation} and~\ref{supp:fig:missing_ablation} compare the three structure-supervision variants on test trajectory G08/P1. In the displayed density region, the full model reduces mean absolute error from 0.0548 to 0.0482 for SR and from 0.1757 to 0.1409 for FC$_s$, relative to training without structure supervision. For FC$_c$, the observed $P,N$ fields condition the shared network's recovery of $W,\Phi_{\mathrm{eff}}$. The full model has the lowest MAE in both displayed missing-field crops. For effective potential, its MAE is 0.0576, compared with 0.0822 without structure supervision and 0.0815 without region balancing.

\twocolumn[{
\subsection{Peak amplitude, recall, and spatial coverage}
\label{supp:area_fraction}
\begin{minipage}{\textwidth}
\begin{suppanchoredtable}
\caption{Training-derived bins and H16 test statistics. Peak values use common normalized fluctuation units. Area fractions are percentages of the SOL. The tails remain open, and boundaries shown here are rounded for display.}
\label{supp:tab:groups}
\setlength{\tabcolsep}{6pt}
\begin{tabular}{lrrrrrr}
\toprule
\multicolumn{7}{l}{\textbf{(a) Strength groups: mean predicted peak within each true region}} \\
Group & Peak interval & Objects & True peak & Full & No structure & No balancing \\
\midrule
1 & $<1.30299$ & 155 & 1.1365 & 0.5148 & 0.4446 & 0.5053 \\
2 & $[1.30299,1.63932)$ & 203 & 1.4627 & 0.7443 & 0.6110 & 0.6660 \\
3 & $[1.63932,2.08294)$ & 169 & 1.8449 & 0.8577 & 0.6991 & 0.8132 \\
4 & $[2.08294,2.66899)$ & 216 & 2.3651 & 1.0410 & 1.0270 & 1.0780 \\
5 & $\geq2.66899$ & 177 & 3.1433 & 1.4839 & 1.3451 & 1.3013 \\
\midrule\multicolumn{7}{l}{\textbf{(b) Area-fraction groups: pooled structure recall (\%)}} \\
Group & Area interval & Windows & Objects & Full & No structure & No balancing \\
\midrule
1 & $<2.01492$ & 18 & 92 & 48.91 & 35.87 & 36.96 \\
2 & $[2.01492,3.76740)$ & 18 & 192 & 45.31 & 40.62 & 38.54 \\
3 & $[3.76740,10.60486)$ & 23 & 397 & 47.61 & 45.34 & 45.59 \\
4 & $\geq10.60486$ & 13 & 239 & 48.12 & 43.51 & 45.61 \\
\bottomrule
\end{tabular}
\end{suppanchoredtable}
\end{minipage}
\vspace{1em}
}]
Table~\ref{supp:tab:groups} groups true structures by peak strength and test windows by area fraction at H16. For each window, the area fraction is the percentage of all SOL pixels in the target frame satisfying $N'_Z\geq\tau$. The bin boundaries come from 288 target frames across the 96 training trajectories. The peak boundaries are quintiles of the 3,898 true structure peaks, and the area-fraction boundaries are quartiles across frames. The first and last bins extend to the tails.

For true structure $A_j$, the true and predicted peaks are
\begin{equation}
v_j=\max_{i\in A_j}N'_Z(i),\qquad
\widehat v_j=\max_{i\in A_j}\widehat N'_Z(i).
\end{equation}
Within each strength group, both means include every true structure, including misses. A displaced peak can fall outside the true region and reduce $\widehat v_j$. Across all five groups, the full model gives a higher mean response within the true region than the model trained without structure supervision. Both models underestimate the true mean peak in every group.

To compute recall in each area-fraction group, we pool the matched and true object counts across its windows. The 36-pixel minimum applies to the connected components counted as objects. Together the four groups contain 920 true structures. The largest recall gain occurs in the sparsest windows, where less than $2.015\%$ of the SOL reaches the threshold. The full model recovers $48.91\%$ of structures there, compared with $35.87\%$ without structure supervision and $36.96\%$ without region balancing. In the amplitude loss, separate region means give the structure region half the weight even when it occupies few pixels.

\newpage
For true structures detected by both the full model and the model without structure supervision, the mean absolute error of the matched peaks is lower for the full model at H8 (0.2285 vs.\ 0.3140) and H32 (0.5630 vs.\ 0.7033), but slightly higher at H16 (0.4754 vs.\ 0.4566).

\begin{figure*}[t]
\centering
\includegraphics[width=\textwidth]{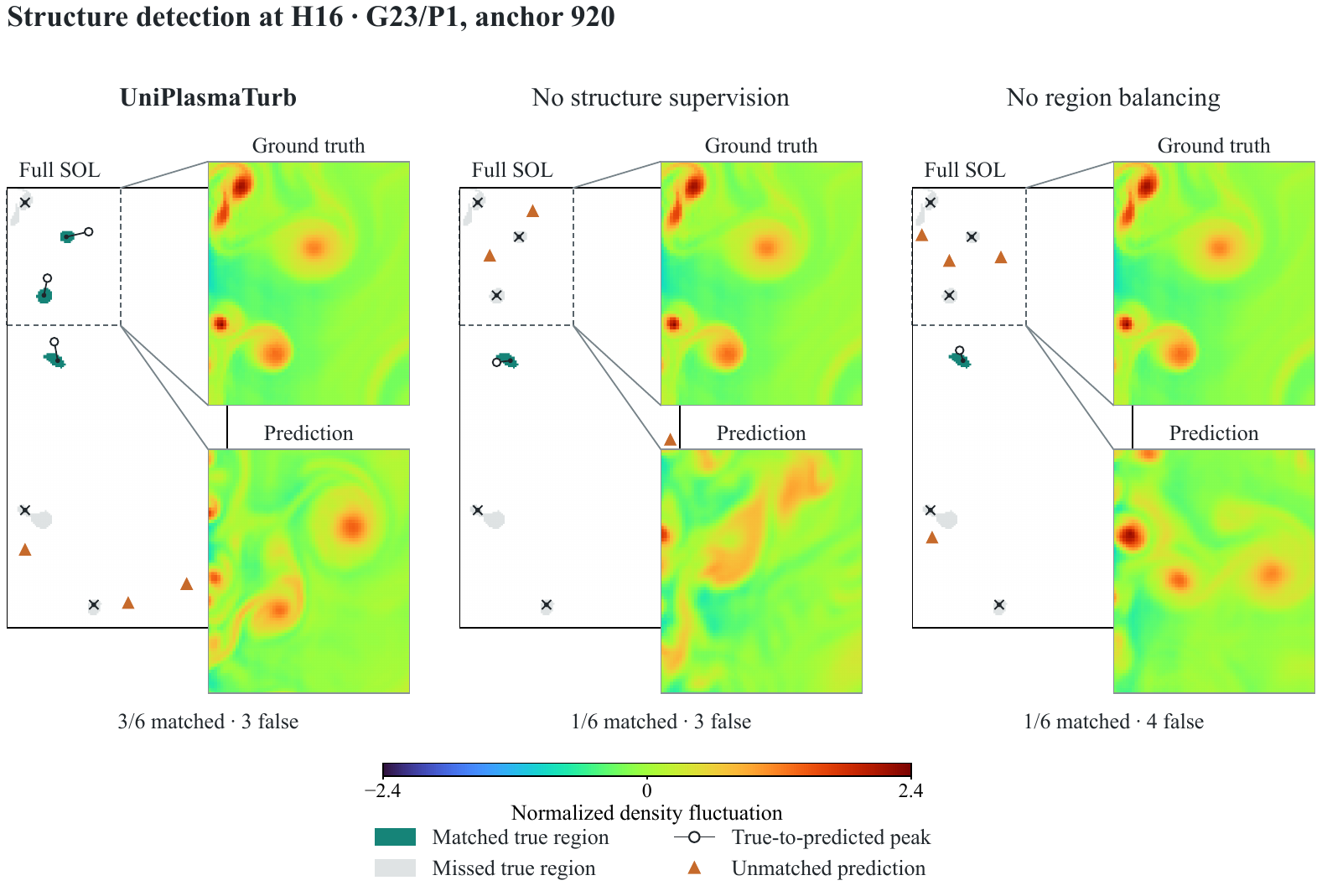}
\caption{\textbf{Recovered and missed density structures at H16.} Test trajectory G23/P1, anchor 920. Full-SOL maps mark matched and missed true regions. Lines connect true peaks to matched predicted peaks (circles), crosses mark misses, and triangles mark unmatched predictions. Gray connectors link the dashed boxes to enlarged field views on a shared linear color scale. Matching uses the 16-pixel periodic distance criterion in Sec.~\ref{supp:protocol}.}
\label{supp:fig:forecast_matches}
\end{figure*}

Figure~\ref{supp:fig:forecast_matches} illustrates object matching in a sparse H16 window. The full model recovers three of six true structures, compared with one for each variant. The connecting segments show the remaining peak displacements, and the unmatched-prediction counts are three, three, and four, respectively.

\paragraph{Spatial coverage across scales.}
The fractions skill score (FSS)~\cite{roberts2008scale} measures how closely the predicted and true fluctuations cover the same neighborhoods. For each $k\times k$ neighborhood in the SOL, let $\hat f_k$ and $f_k$ denote the predicted and true fractions of pixels at or above $\tau$. Neighborhoods wrap in $y$ and must lie entirely within the SOL in $x$. We compute $\mathrm{FSS}_k=1-\sum(\hat f_k-f_k)^2/\sum(\hat f_k^2+f_k^2)$, pooling both sums over neighborhood centers and all 72 test windows.

\begin{figure*}[t]
\centering
\includegraphics[width=\textwidth]{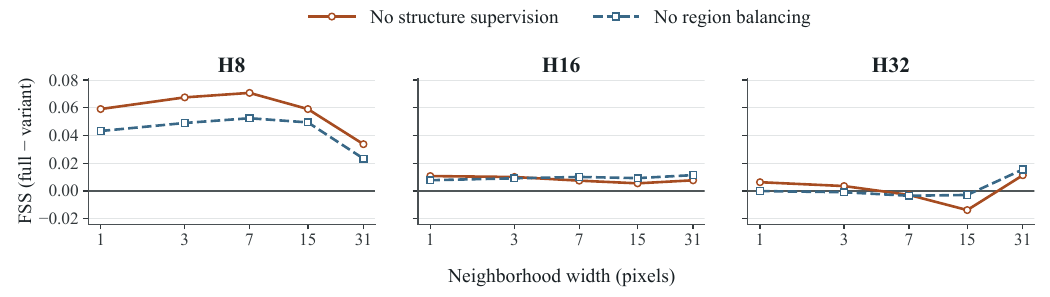}
\caption{\textbf{Spatial coverage of strong density fluctuations across forecast horizons.} Each curve shows the full model's FSS minus that of the indicated variant. Positive values indicate closer agreement with the true local coverage.}
\label{supp:fig:coverage}
\end{figure*}

At H8, the full model gives closer coverage than both variants at all five scales (Fig.~\ref{supp:fig:coverage}). At the $7\times7$ scale used by the local occupancy loss, FSS reaches 0.536, compared with 0.465 without structure supervision and 0.483 without region balancing. The advantage persists at H16 but is smaller. At H32, both variants score higher at widths of 7 and 15 pixels, while the full model scores highest at 31 pixels.

\twocolumn
\suppresetcolumns
\subsection{Effect of distinct historical states at fixed weights}
\label{supp:history}
The history analysis keeps the epoch-32 weights and all four input slots fixed. If the available window is $(U_{t-3},U_{t-2},U_{t-1},U_t)$, the three inputs are
\begin{equation}
\begin{aligned}
\mathcal H^{(1)}&=(U_t,U_t,U_t,U_t),\\
\mathcal H^{(2)}&=(U_{t-1},U_{t-1},U_{t-1},U_t),\\
\mathcal H^{(4)}&=(U_{t-3},U_{t-2},U_{t-1},U_t).
\end{aligned}
\end{equation}
At each forecasting step, the model applies the same rule to the current window. The latest state is the last observation at the first step and the previous prediction at every subsequent step.

For FC$_c$, using one, two, and four distinct states gives $W$ MSE of $0.08251$, $0.03177$, and $0.03011$, respectively. The corresponding effective-potential errors are $0.28758$, $0.01588$, and $0.01073$. Temporal variation in the observed $P,N$ fields thus improves inference of the unobserved fields.

The same history also improves short-horizon localization. Using four states instead of one reduces the mean localization error in Eq.~\eqref{supp:eq:localization} from $12.00$ to $9.68$ pixels at H8 and from $13.91$ to $12.90$ at H16. At H32 the errors with two and four states are similar, at $13.48$ and $13.56$ pixels, respectively.

\subsection{Error versus distance from the periodic seam}
\label{supp:seam}
For row index $y=0,\ldots,255$, define the folded seam distance $d_y=\min(y,255-y)$. Rows at the same distance from the two ends of the periodic axis are combined into eight-pixel bins,
\begin{equation}
B_b=\{y:8b\leq d_y<8(b+1)\},\qquad b=0,\ldots,15.
\end{equation}
At each forecast horizon, squared errors in common units are averaged over the four fields, all radial positions, rows in the bin, and the 72 windows. The bin centers plotted in Fig.~\ref{fig:periodic_boundary_analysis} are $3.5,11.5,\ldots,123.5$ pixels. Equal bin sizes make the average over bins equal to the corresponding complete-field MSE.

Periodic convolution reduces one-step error near the seam. In the nearest eight pixels, H1 MSE is $0.01966$ with periodic padding and $0.04456$ with replicated padding. At H8 the advantage extends into the interior. By H32 the gap narrows, and replicated padding gives lower error in some distance bins.

\section{Scope And Limitations}
\label{supp:limitations}
The dataset varies the initial perturbation in two-dimensional local simulations with fixed backgrounds and boundaries. Extending the simulations to other plasma regimes would test how well the model transfers across physical conditions. The reconstruction tasks use exact coarse or partial observations. Applying the framework to experimental diagnostics requires accounting for measurement uncertainty in the output projection.

As forecasts extend from H8 to H32, the full model recovers a smaller fraction of the true density structures and its mean localization error increases. Future work will track individual structures across frames and assess how forecast errors affect estimates of particle transport.

\end{document}